\documentclass[aps,pra,reprint,superscriptaddress,showkeys,nofootinbib]{revtex4-2}

\usepackage[dvipsnames]{xcolor}
\usepackage[colorlinks=true,linkcolor=OrangeRed,citecolor=Cerulean,urlcolor=ProcessBlue]{hyperref}
\usepackage{amsmath}
\usepackage{amsfonts}
\usepackage{amssymb}
\usepackage{bbm}
\usepackage{dsfont}
\usepackage{mathrsfs}
\usepackage{array}
\usepackage{dcolumn}
\usepackage{url}
\usepackage{textcomp}
\usepackage{graphicx}
\usepackage{color}
\usepackage{braket}
\usepackage[mathscr]{euscript}
\usepackage{bbm}
\usepackage{enumerate}
\usepackage{appendix}
\usepackage{lipsum}
\usepackage[10pt]{moresize}
\usepackage{float}

\def\alp{\alpha}
\def\bt{\beta}
\def\sg{\sigma}

\def\om{\omega}
\def\Om{\Omega}

\def\lo{\lambda}

\def\Dt{\Delta}
\def\dt{\delta}
\def\gm{\gamma}

\def\ep{\epsilon}

\def\8{\phi}
\def\7{\Phi}
\def\9{\psi}

\def\fr{\frac}

\def\td{\tilde}

\def\b{\bra}
\def\B{\Bra}
\def\kt{\ket}
\def\K{\Ket}
\def\bk{\braket}
\def\BK{\Braket}
\def\h{\hat}

\def\ra{\rightarrow}

\def\V{\vec}

\def\tens{\otimes}
\def\dsum{\oplus}

\def\Cm{\mathcal{C}}
\def\Dm{\mathcal{D}}
\def\Em{\mathcal{E}}

\def\Hm{\mathcal{H}}

\def\Jm{\mathcal{J}}

\def\Lm{\mathcal{L}}

\def\Oe{\mathcal{O}}
\def\Pm{\mathcal{P}}
\def\Rm{\mathcal{R}}

\def\Tm{\mathcal{T}}

\def\Zm{\mathcal{Z}}
\def\iD{\mathbbm{1}}

\def\RN{\mathbbm{R}}

\def\3{\big}
\def\4{\Big}
\def\5{\Bigg}
\definecolor{ogreen}{rgb}{0.2, 0.5, 0.1}
\definecolor{Orange}{rgb}{1, 0.6, 0.4}

\def\ti{\textit}

\def\tt{\text}
\newcommand{\tus}[1]{\underline{\smash{#1}}}
\def\sz{\small}
\def\fsz{\footnotesize}
\newcommand{\tsz}[1]{\text{\small #1}}

\def\aref{\autoref}
\newcommand{\appref}[1]{\hyperref[#1]{Appendix~\ref{#1}}}
\newcommand{\apprefNull}[1]{\hyperref[#1]{Appendix}}
\usepackage[english]{babel}
\addto\extrasenglish{%
}

\addto\captionsenglish{}

\def\vh{\vspace{0.5mm}}
\def\vo{\vspace{1mm}}
\def\vt{\vspace{2mm}}
\def\vth{\vspace{3mm}}
\def\vfo{\vspace{4mm}}
\def\vf{\vspace{5mm}}

\def\vTo{\vspace{-1mm}}
\def\vTt{\vspace{-2mm}}
\def\vTh{\vspace{-3mm}}
\def\vFo{\vspace{-4mm}}
\def\vEq{\vspace{-2mm}}
\def\hs{\kern 0.16667em}
\def\Hs{\kern 0.5em}
\def\hsh{\kern 0.1em}
\def\hshh{\kern 0.07em}
\def\hsn{\kern -0.3em}
\def\hsN{\kern -0.6em}
\def\hseN{\kern -0.55em}
\def\hsnh{\kern -0.15em}
\def\hsnhh{\kern -0.1em}

\newcommand{\rt}[1]{\raisetag{#1\baselineskip}}

\newcommand*{\Scale}[2][4]{\scalebox{#1}{$#2$}}%

\newcommand{\SDrop}[2]{\fontdimen16\textfont2=#1
\fontdimen17\textfont2=#1 \tt{$#2$} \fontdimen16\textfont2=2pt
\fontdimen17\textfont2=2pt}
\newcommand{\SRaise}[2]{\fontdimen13\textfont2=#1
\fontdimen14\textfont2=#1 \fontdimen15\textfont2=#1 \tt{$#2$} \fontdimen13\textfont2=2pt
\fontdimen14\textfont2=2pt \fontdimen15\textfont2=2pt}

\usepackage{fancyhdr}
\numberwithin{equation}{section}

\def\tr{\text{Tr}}
\def\var{\text{Var}}

\def\methodAT{\fsz TAME\sz}
\def\methodATO{\fsz TAME \sz}
\def\methodT{\sz TAME\normalsize}
\def\methodTO{\sz TAME \normalsize}

\def\benchMT{\sz ETA\normalsize}
\def\benchMTO{\sz ETA \normalsize}

\def\vIT{\sz ETA-Variant I\normalsize}
\def\vITO{\sz ETA-Variant I \normalsize}
\def\vIIT{\sz ETA-Variant II\normalsize}
\def\vIITO{\sz ETA-Variant II \normalsize}

\usepackage[all]{hypcap}

\usepackage[shortlabels]{enumitem}

\begin{document}

\title{Benchmarking Quantum Simulators}
\date{\today}

\author{Andrew Shaw}
\email[Electronic Address: ]{ashaw12@umd.edu}
\affiliation{University of Maryland, College Park, MD 20742, USA}

\begin{abstract}

\ti{Time-Averaged Mixed-state Equivalence} (\methodAT) is used to benchmark quantum simulators with classical computing resources. The classical computation is feasible even if direct computation of the real-time dynamics is prohibitively costly.
 
\end{abstract}

\maketitle

%%-------------Section I--------------------%%

\section{Time-Averaged Mixed-state Equivalence}

\ti{Time-Averaged Mixed-state Equivalence} (\methodT) relates the real-time dynamics of \ti{pure quantum states} and the expectation values of \ti{mixed quantum states}.

\subsection{Time-Averaged Dynamics}

An observable \tsz{$\h O$} undergoes time evolution governed by a Hamiltonian \tsz{$\h H$}: 

\vEq

\begin{align}
\h H&=\sum_{\alp} \Em_{\alp} \kt{\alp}\hsnh \b{\alp},\\ 
\h O&=\sum_{\alp,\bt} \Oe_{\alp,\bt} \kt{\alp}\hsnh \b{\bt}
\end{align}

The real-time dynamics of \tsz{$\h O$} are as follows  \cite{tame0}:

\vEq

\begin{equation}
\h O(t)= \hs \SRaise{5pt}{e^{\Scale[0.92]{\hsh i\h H t}}} \hs \h O \hs \SRaise{5pt}{e^{-\Scale[0.92]{i\h H t}}}		
\end{equation}

\vTh

\begin{equation}
\h O(t)=\sum_{\alp,\bt} \SDrop{3pt}{\Oe_{\Scale[0.78]{\alp,\bt}}} \hs \SRaise{6pt}{\Scale[1.05]{e}^{-\Scale[0.92]{it \3(\Em_{\bt}-\Em_{\alp}\3)}}} \kt{\alp}\hsnh \b{\bt}
\end{equation}

The \ti{time-averaged observable} is the following:

\vEq

\begin{equation}
\h \Om=\lim_{t\ra \infty} \fr{1}{t} \int_0^t \hs dt^{'} \hs \h O(t^{'})
\end{equation}

\vTh

\begin{equation}
\h \Om= \sum_{\alp} \Oe_{\alp,\alp} \kt{\alp}\hsnh \b{\alp}
\end{equation}

The \ti{time-averaged expectation value} of a pure state \tsz{$\kt{\9}$} is the following:

\vEq

\begin{equation}
\bk{\9|\hsh \h \Om \hsh|\9}=\sum_{\alp} |c_{\alp}|^2 \hs \Oe_{\alp,\alp}	
\end{equation}

\vTh

\begin{equation}
\kt{\9}=\sum_{\alp} c_{\alp} \kt{\alp}
\end{equation}

\newpage

\subsection{Mixed State Expectation Values}

A mixed quantum state is composed of an ensemble of pure states \tsz{$\kt{\SDrop{3pt}{\9_k}}$}, with observational probabilities \tsz{$\SDrop{3pt}{p_k}$}. Its \ti{density matrix} \cite{densmat0,densmat1} is the following:

\vEq

\begin{align}
\rho &=\sum_k \SDrop{4pt}{p_k} \K{\SDrop{3pt}{\9_k}}\hsnh \b{\SDrop{3pt}{\9_k}}, \\
&=\sum_{\alp,\bt}  \Pm_{\alp,\bt} \kt{\alp}\hsnh \b{\bt}
\end{align}

The expectation value of \tsz{$\h O$} is as follows:

\vEq

\begin{equation}
\bk{\h O}=\tr \4[\rho \hs \h O\hsh \4]
\end{equation}

\vTh

\begin{equation}
\bk{\h O}=\sum_{\alp,\bt} \Pm_{\alp,\bt}\hs \Oe_{\bt,\alp}
\end{equation}

\ti{Orthodox mixed states} commute with the Hamiltonian:

\vEq

\begin{align}
& \ \ \ \ \4[\hsh\td \rho, \h H \hsh\4]=0 \\[0.6em]
&\td \rho=\sum_{\alp}  \td \Pm_{\alp} \kt{\alp}\hsnh \b{\alp}
\end{align}

The \ti{orthodox expectation value} is the following:

\vEq

\begin{equation}
\tr \4[\td \rho \hs \h O\hsh \4]=\sum_{\alp} \td \Pm_{\alp}\hs \Oe_{\alp,\alp}
\end{equation}

Density matrices must be \ti{positive semi-definite} with \tsz{$\tr(\rho)=1$} \cite{densmat2}, which requires the following conditions to be satisfied:

\vEq

\begin{equation}
\sum_{\alp} \td \Pm_{\alp}=1, \ \ \td \Pm_{\alp} \in \RN^{+}
\end{equation}

\begin{figure}
\includegraphics[scale=0.039]{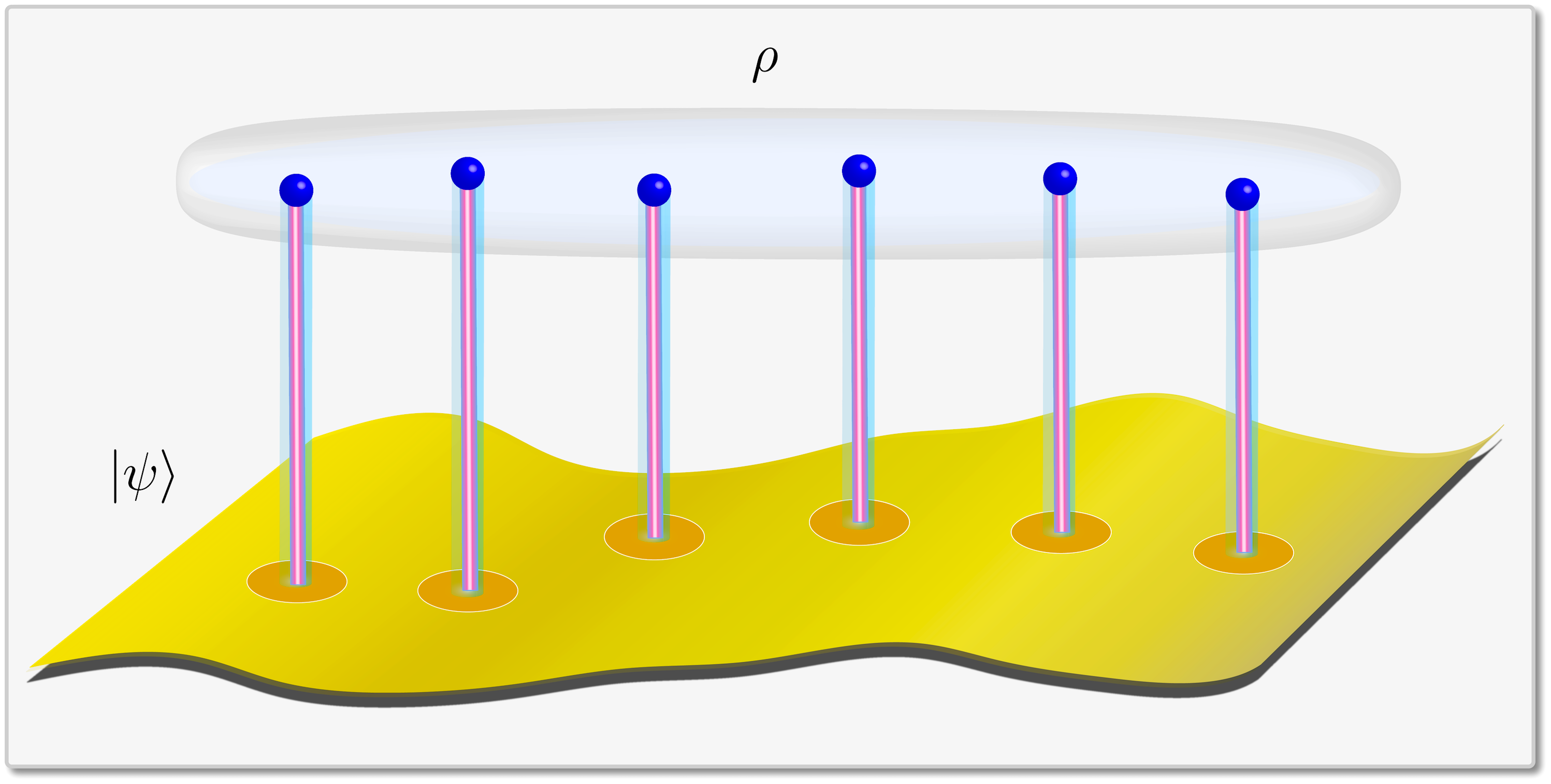}
\caption{ [Upper Panel]: Orthodox mixed states (blue) reside in the space of mixed states (gray). [Lower Panel]: \methodATO relates each orthodox mixed state to a subset of states (golden) in the space of pure states (yellow).}
\label{fig:orthodox}
\end{figure}

The time-averaged expectation values of each pure state are equivalent to the expectation values of an orthodox mixed state. This mapping is not one-to-one, as a continuum of pure states share a corresponding orthodox mixed state (\aref{fig:orthodox}).

\subsection{Coarse-Grained TAME}

\methodTO is coarse-grained by integrating over the pure states and the orthodox mixed states \cite{haar0,haar1,haar2,haar3}.

\newpage
%
%This can be accomplished by ensuring that only states with real expansion coefficients contribute to the integral over pure states:
%
%\begin{align}
%\int d\9\ra \int &d\9 \ \Km(\9) \\
%\Km(\9)=\prod_{\alp}\5[\dt\4\{ &\tt{Arg}\3[\bk{\alp|\9}\3]\4\}\5]
%\end{align}

\vo

The \ti{coarse-grained time-averaged expectation value} is obtained by integrating over all pure states:

\vEq

\begin{align}
\bk{\h \Om}_c&=\int d\9 \ \bk{\9|\hsh\h \Om \hsh|\9} \\
&=\int d\9 \ \sum_{\alp} 	{|c_\alp(\9)|}^2 \hs \Oe_{\alp,\alp}\\
&=\sum_{\alp}\hs \Oe_{\alp,\alp} \int d\9 \ 	\hs {|c_\alp(\9)|}^2  \\
&=\tr\3[\h O\3]\cdot \fr{1}{\tt{dim}\{\h H\}}
\end{align}

\vt

The \ti{coarse-grained orthodox expectation value} is obtained by integrating over all orthodox mixed states:

\vEq

\begin{align}
\bk{\h O}_c &=\int d\td \rho \ \hsh \tr \4[\td \rho \hs \h O \4]  \notag\\
&=\int d\td \rho \ \sum_{\alp} \td \Pm_{\alp}\hs \Oe_{\alp,\alp} \\
&= \sum_{\alp} \Oe_{\alp,\alp} \int d \td \rho \ \td \Pm_{\alp} \\
&=\tr\3[\h O\3] \cdot \fr{1}{\tt{dim}\{\h H\}}
\end{align}

Integrating over the pure states and the orthodox mixed states yields an identical quantity.

%%-------------Section II--------------------%%

\section{Quantum Simulation Benchmark}

Benchmarking the output of a \ti{quantum simulator} \cite{feynmansupremacySim,maninsupremacySim} can be accomplished with \methodT. This requires determining the expectation values of orthodox mixed states using \ti{classical computing resources} \cite{ClassComp0}.

\subsection{Orthodox Mixed State Computation}

Orthodox mixed states can be expressed by applying positive-real functions \tsz{$f(x)$} to the Hamiltonian:

\vEq

\begin{align}
\td \rho(f)&=\fr{f(\h H)}{\tr\4[f(\h H)\4]} \\
&=\fr{1}{\Zm} \hs  f(\h H)
\end{align}

A positive-real function can be expressed as follows:

\vEq

\begin{align}
f(x)&= \int_{-\infty}^{\infty} dt \ \dt(x-t) \ f(t) \\
&=\lim_{\sg\ra 0}\hs 	\fr{1}{|\sg| \sqrt{\pi}}\int_{-\infty}^{\infty} dt \ e^{-\4(\Scale[1.2]{\fr{x-t}{\sg}}\4)^2} \hs f(t)
\end{align}

A general orthodox mixed state can be expressed as follows:

\vEq

\begin{equation}
\td \rho(f)=\fr{1}{\Zm}\5[\lim_{\sg\ra 0} \hs \fr{1}{|\sg| \sqrt{\pi}} \int_{-\infty}^{\infty} dt \ e^{-\4(\Scale[1.2]{\fr{\h H-t}{\sg}}\4)^2} \hs f(t)\5]
\end{equation}

As such, it is sufficient to consider \ti{gaussian orthodox mixed states}:
\vEq

\begin{equation}
\td \rho_g(\ep,\sg) =\fr{1}{\Zm} \  e^{-\4(\Scale[1.2]{\fr{\h H-\ep}{\sg}}\4)^2}
\end{equation}

These can be re-expressed in terms of a Hermitian operator \tsz{$\h K$} and a real parameter \tsz{$\tau$}:

\begin{equation}
\td \rho_g(\ep,\sg)=\fr{1}{\Zm} \ \SRaise{4pt}{e^{-\Scale[0.92]{\h K \tau}}}
\end{equation}

The expectation value of \tsz{$\h O$} is as follows:

\vEq

\begin{align}
\bk{\h O}&=\fr{1}{\Zm}\tr\4[\SRaise{4pt}{e^{-\Scale[0.92]{\h K \tau}}}\h O \hsh \4] \\
&=\fr{1}{\Zm} \sum_{a} \bk{a|\hs \SRaise{4pt}{e^{-\Scale[0.92]{\h K \tau}}}\hsh \h O \hs |a}		
\end{align}

The expectation value can be classically computed in a manner similar to the \ti{path integral} \cite{pathint0}. This is accomplished by discretizing \tsz{$\SRaise{4pt}{e^{-\Scale[0.92]{\h K \tau}}}$}:

\vEq

\begin{equation}
\SRaise{4pt}{e^{-\Scale[0.92]{\h K \tau}}}\ra \SRaise{4pt}{e^{-\Scale[0.85]{\h K \Dt}}}\hsh \SRaise{4pt}{e^{-\Scale[0.85]{\h K \Dt}}}
\hsh \hdots \hsh \SRaise{4pt}{e^{-\Scale[0.85]{\h K \Dt}}}
\end{equation}

The identity is inserted between the matrix exponentials:

\begin{equation}
\begin{split}
\SRaise{4pt}{e^{-\Scale[0.92]{\h K \tau}}}=&\sum_{a_1}\hdots \sum_{a_N} \ \kt{a_1}\hsnh\b{a_1} \SRaise{4pt}{e^{-\Scale[0.85]{\h K \Dt}}} \kt{a_2}\hsnh \b{a_2}\SRaise{4pt}{e^{-\Scale[0.85]{\h K \Dt}}}\kt{a_3}\hsnh\b{a_3}	\\
& \qquad \hdots \kt{a_{N-1}}\hsnh \b{a_{N-1}}\SRaise{4pt}{e^{-\Scale[0.85]{\h K \Dt}}} \kt{a_N}\hsnh \b{a_N}
\end{split}\rt{1}
\end{equation}

\newpage

This can be re-expressed as a sum over \ti{configurations} \tsz{$\V a=\{a_1,\cdots,a_N\}$}:

\vEq

\begin{align}
\SRaise{4pt}{e^{-\Scale[0.92]{\h K \tau}}} &=\sum_{\V a} S(\V a) \kt{a_1}\hsnh \b{a_N} \\[0.6 em]
S(\V a)=\b{a_1}& \SRaise{4pt}{e^{-\Scale[0.85]{\h K \Dt}}}  \kt{a_2}\hdots \b{a_{N-1}}\SRaise{4pt}{e^{-\Scale[0.85]{\h K \Dt}}} \kt{a_N}
\end{align}

\vt

The expectation value of \tsz{$\h O$} is the following:

\vTh

\begin{align}
\bk{\h O}&=\fr{1}{\Zm} \sum_{a} \5[\sum_{\V a} \b{a} S(\V a) \kt{a_1}\hsnh \b{a_N} \h O \kt{a}	\5]\\
&=\fr{1}{\Zm} \sum_{\V a} S(\V a) \b{a_N}\h O\kt{a_1} \\
%&=\fr{1}{\Zm} \sum_{\V a} \3|S(\V a)\3| \ e^{i\8(\V a)} \b{a_N}\h O\kt{a_1}  \\
&=\fr{1}{\Zm} \sum_{\V a} \3| S(\V a) \3| \hs O(\V a)
\end{align}

Classical computing resources can be used to sample from the configurations with the following probability:\footnote{\ti{Monte-Carlo techniques} can be used to sample from a target probability distribution (\apprefNull{appendix:AMonteCarlo}).}

\begin{equation}
P(\V a_{\tt{\tiny samp}} \in \V a_j\ )=	\fr{1}{\xi}\fr{\3|S(\V a_j)\3|}{\Zm}
\end{equation}

%\begin{equation}
%\fr{P(\V a_{\tt{\tiny samp}} \in \V a_j\ )}{P(\V a_{\tt{\tiny samp}} \in \V a_k\ )}=	\fr{\3|S(\V a_j)\3|}{\3|S(\V a_k)\3|}
%\end{equation}

The sampled configurations can be used to estimate \tsz{$\bk{ \h O}$} as follows:

\vEq

\begin{equation}
\bk{\h O} \approx \fr{\xi}{N_{\tt{samp}}} \sum_{k=1}^{N_{\tt{samp}}}	O\4(\V a^{(k)}_{\tt{\tiny samp}}\4)
\end{equation}

\subsection{Direct Benchmark}

Classical computing resources can be used to compute the orthodox expectation value of \tsz{$\td \rho(f)$}:

\vEq

\begin{align}
\td \rho(f)&=\fr{1}{\Zm} \hs  f(\h H) \\
&=\fr{1}{\Zm}\sum_{\alp} f(\Em_{\alp}) \kt{\alp}\hsnh \b{\alp}
\end{align}

The corresponding \ti{direct benchmarking states} take the following form:

\vEq 
 
\begin{equation}
\kt{\9(f)}=\sum_{\alp} e^{\Scale[0.95]{i\8_{\alp}}} \sqrt{\fr{f(\Em_{\alp})}{\Zm}} \kt{\alp}
\end{equation}

The direct benchmarking states satisfy the following property:

\vEq

\begin{equation}
\bk{\9(f)|\hs  \h \Om \hs 	|\9(f)}=\tr\4[\td \rho(f) \h O\4]
\end{equation}

A quantum simulator generates dynamics governed by the \ti{simulation Hamiltonian} \tsz{$\h H_s$}:

\begin{equation}
\h H_s=\sum_{\Scale[0.8]{\alp^s}} \Em^s_{\alp} \hsh \kt{\SRaise{4.8pt}{\alp^s}}\hsnh \b{\SRaise{4.8pt}{\alp^s}}	
\end{equation}

The quantum simulator provides access to the \ti{simulated observable}:

\begin{equation}
\h O_s(t)= \hs \SRaise{5pt}{e^{\Scale[0.92]{\hsh i\h H_s t}}} \hs \h O \hs \SRaise{5pt}{e^{-\Scale[0.92]{i\h H_s t}}}	
\end{equation}	

The \ti{simulated time-averaged observable} is as follows:

\vTt

\begin{equation}
\h \Om_s=\lim_{t\ra \infty} \fr{1}{t} \int_0^t \hs dt^{'} \hsh \h O_s(t^{'})
\end{equation}

If the quantum simulator exactly reproduces the target dynamics, the \ti{simulated time-averaged expectation value} will recover the orthodox expectation value for all direct benchmarking states:

\begin{align}
\lim_{\h H_{\Scale[0.7]{s}}\ra \h H} \bk{\9(f)|\hs \h \Om_s \hs |\9(f)} \hs =	\hs \tr\4[\td \rho(f) \h O\4], \ \forall \kt{\9(f)}
\end{align}

\subsection{Coarse-Grained Benchmark}

An analogous procedure can be performed by coarse-graining \methodTO over a portion of the Hilbert space.

\subsubsection{Projecting Coarse-Grained \methodT}

The Hamiltonian can be written in the following form:

\vEq

\begin{equation}
\h H=\h H_P \dsum \h H_Q	
\end{equation}

The Hilbert space of \tsz{$\h H$} can be expressed as follows:

\vEq

\begin{equation}
\Hm=\Hm_P\dsum \Hm_Q	
\end{equation}

The Hilbert subspace \tsz{$\Hm_Q$} is spanned by the states \tsz{$\kt{\SDrop{3pt}{Q_{\hsnhh i}}}$}:

\vEq

\begin{equation}
\Hm_Q=\overline{\tt{span}}\3\{\hsh \kt{Q_1},\hsh \kt{Q_2}, 
\ \Scale[1.2]{\hdots} \ \3\}
\end{equation}

\vt

The \ti{projection operator} \cite{projec0} onto \tsz{$\Hm_P$} is the following:

\begin{equation}
\h P=\h \iD	-\sum_{i} \kt{Q_i}\hsnh \b{Q_i}
\end{equation}

To coarse-grain \methodTO on \tsz{$\Hm_P$}, pure states and orthodox mixed states with support on \tsz{$\Hm_Q$} are excluded from the integral.

\newpage

To enforce this exclusion, the integral over pure states is modified as follows:

\vEq

\begin{align}
& \ \ \ \int d\9\ \ra \int d\9 \ \Scale[1.1]{\Jm}\3(\Scale[0.85]{\h P},\9 \hshh \3) \\[0.4em]
&\Scale[1.1]{\Jm}(\Scale[0.85]{\h P},\9 \hshh)=\dt\4\{1-\bk{\9|\hsh \h P \hsh |\9}\4\}
\end{align}

\vo

Likewise, the integral over orthodox mixed states is modified as follows:

\vEq

\begin{align}
& \ \ \ \int d\td \rho \ra \int d \td \rho \ \hs	\Scale[1.1]{\Rm}\3(\Scale[0.85]{\h P},\td \rho \hsh\3) \\[0.4em]
&\Scale[1.1]{\Rm}\3(\Scale[0.85]{\h P},\td \rho \hsh \3) =\dt \4\{1-\tr\3[\h P \hs \td \rho \hs \3]\4\}
\end{align}

\vo

The \ti{projected coarse-grained time-averaged expectation value} is the following:

\vEq

\begin{align}
\bk{\h \Om}_{c,P}&=\int d\9 \ \hs \Scale[1.1]{\Jm}\3(\Scale[0.85]{\h P},\9 \hshh\3) \hsh  \bk{\9|\hs \h \Om\hs |\9} \\
%&=\sum_{\alp} \Oe_{\alp,\alp} \int d\9 \ \Jm(\9) \hs |c_{\alp}(\9)|^2\\
&=\tr\3[\h P\hs \h O \hsh \3]\cdot \fr{1}{\tt{dim}\{\h H_P\}}
\end{align}

\vo

The \ti{projected coarse-grained orthodox expectation value} is the following:

\vEq 

\begin{align}
\bk{\h O}_{c,P}&=	\int d\td \rho \ \hs \Scale[1.1]{\Rm}\3(\Scale[0.85]{\h P},\td \rho \hsh \3) \hs \hsh \tr\4[\td \rho \hs \h O\4]\\
&=\tr\3[\h P \hs \h O \hsh \3]\cdot \fr{1}{\tt{dim}\{\h H_P\}}
\end{align}

\vt

Integrating over compatible subsets of the pure states and orthodox mixed states yields an identical quantity.

\subsubsection{Simulation Benchmark}

A \ti{subspace mapping} \tsz{$\Lm$} uses a Hamiltonian to specify a Hilbert subspace \tsz{$\Hm_{P}$} with projection operator \tsz{$\h P$}:

\vEq

\begin{align}
&\Lm(\h H)=\Hm_{P} \\[0.4em]
& \hs \3[\h P,\h H\3]=0	
\end{align}

\vo

\ti{Benchmark orthodox mixed states} \tsz{$\td\rho_b$} have support solely on \tsz{$\Hm_P$}:

\vEq

\begin{equation} 
\td \rho_b \in \4\{\Hm\tens\SRaise{6pt}{\Hm^*} \hsnh \4\}_P, \ \forall \ \td \rho_b
\end{equation}

\vo

The projected coarse-grained orthodox expectation value is approximated by sampling from the benchmark orthodox mixed states with a uniform probability:

\begin{equation}\label{eq:standcomp}
\bk{\h O}_{c,P}\approx	\fr{1}{N_{\tt{samp}}}\sum_{i=1}^{N_{\tt{samp}}} \tr\4[\td \rho_b^{(i)} \h O \4]
\end{equation}

The subspace mapping is applied to the simulation Hamiltonian to specify a Hilbert subspace \tsz{$\Hm_{\SDrop{0pt}{P_{\Scale[1]{s}}}}$} with projection operator \tsz{$\h P_s$}:

\vTh

\begin{align}
&\Lm(\h H_s)=\Hm_{\SDrop{0pt}{P_{\Scale[1]{s}}}} \\[0.4em]
& \hs \3[\h P_s,\h H_s\3]=0	
\end{align}

\vo

\ti{Coarse-grained benchmarking states} \tsz{$\kt{\9_b}$} have support solely on \tsz{$\Hm_{\SDrop{0pt}{P_{\Scale[1.15]{s}}}}$}:

\vFo

\begin{equation}
\kt{\9_b} \in \Hm_{\SDrop{0pt}{P_{\Scale[1]{s}}}}, \ \forall \hs \kt{\9_b}
\end{equation}

\vo

The quantum simulator can be used to approximate the \ti{projected coarse-grained simulated time-averaged expectation value}:

\vTh

\begin{equation}
\bk{\h \Om_s}_{c,\SDrop{0pt}{P_{\Scale[1]{s}}}}=\int d\9 \ \hs \Scale[1.1]{\Jm}\3(\Scale[0.85]{\h P_{\Scale[0.88]{s}}},\9 \hshh \3) \hsh  \bk{\9|\hs \h \Om_s\hs |\9}
\end{equation}

\begin{figure}
\includegraphics[scale=0.040]{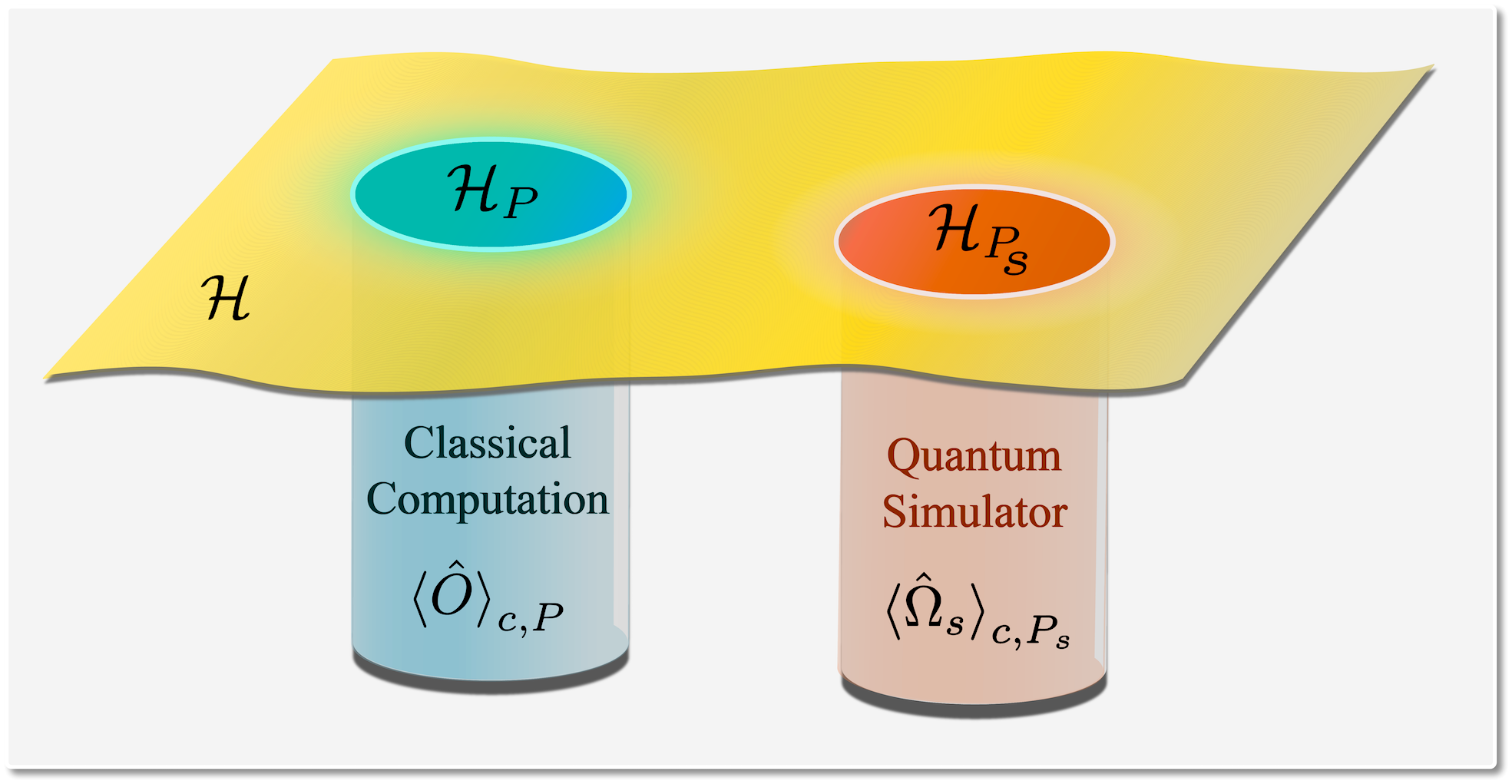}
\caption{ [Upper Panel]: $\Hm_P$ (blue) and $\Hm_{\SDrop{0pt}{P_{\Scale[1.15]{s}}}}$ (orange) are subspaces of $\Hm$ (yellow) that are specified by $\h H$ and $\h H_s$. [Lower Panel]: Classical computation and quantum simulation are used to probe the equivalence of $\Hm_P$ and $\Hm_{\SDrop{0pt}{P_{\Scale[1.15]{s}}}}$.}
\label{fig:subspace}
\end{figure}

\vo

This is accomplished by sampling from the coarse-grained benchmarking states with a uniform probability:

\vEq

\begin{equation}
\bk{\h \Om_s}_{c,\SDrop{0pt}{P_{\Scale[1]{s}}}} \approx \fr{1}{N_{\tt{samp}}}\sum_{i=1}^{N_{\tt{samp}}} \b{\9^{(i)}_b}\h \Om_s\kt{\9^{(i)}_b}	
\end{equation}
 
\vt
 
 If \tsz{$\Hm_P=\Hm_{\SDrop{0pt}{P_{\Scale[1.15]{s}}}}$}, the projected coarse-grained orthodox expectation value will equal the projected coarse-grained simulated time-averaged expectation value (\aref{fig:subspace}).
 
 \vTo
  
%%-------------Section III--------------------%%

\section{Energy-Window Time-Averaging}

\vTo

\ti{Energy-window Time-Averaging} (\benchMT) is a coarse-grained benchmarking procedure. Coarse-grained benchmarking can be represented schematically in three stages:

\vo

\begin{enumerate}[I.]
\item \tus{Projection}: Establish a subspace mapping.
\item \tus{Standardization}: Sample benchmark orthodox mixed states to establish a \ti{standard of comparison}.
\item \tus{Arbitration}: Sample coarse-grained benchmarking states to establish a \ti{simulation diagnostic}.
\end{enumerate}

\subsection{Projection}

Both \vITO and \vIITO require an \ti{energy-window} \tsz{$\Dt_{\ep}$} to perform projection:

\vEq

\begin{equation}
\ep_{\tt{min}}< \Em < \ep_{\tt{max}}, \ \forall \Em \in \Dt_{\ep}
\end{equation}

\subsubsection{Variant I}

In \vIT, the subspace mapping uses the expectation value of the Hamiltonian to define a subset \tsz{$\Hm_P^{(I)}\in \Hm$}. States with average energies that fall within the energy-window are members of the subset (\aref{fig:ewindVarI}):

\begin{equation}
\b{\9}\h H\kt{\9} \in \Dt_{\ep},\  \forall  \kt{\9} \in \Hm^{(I)}_{P}
\end{equation}

\subsubsection{Variant II}

\begin{figure}
\includegraphics[scale=0.040]{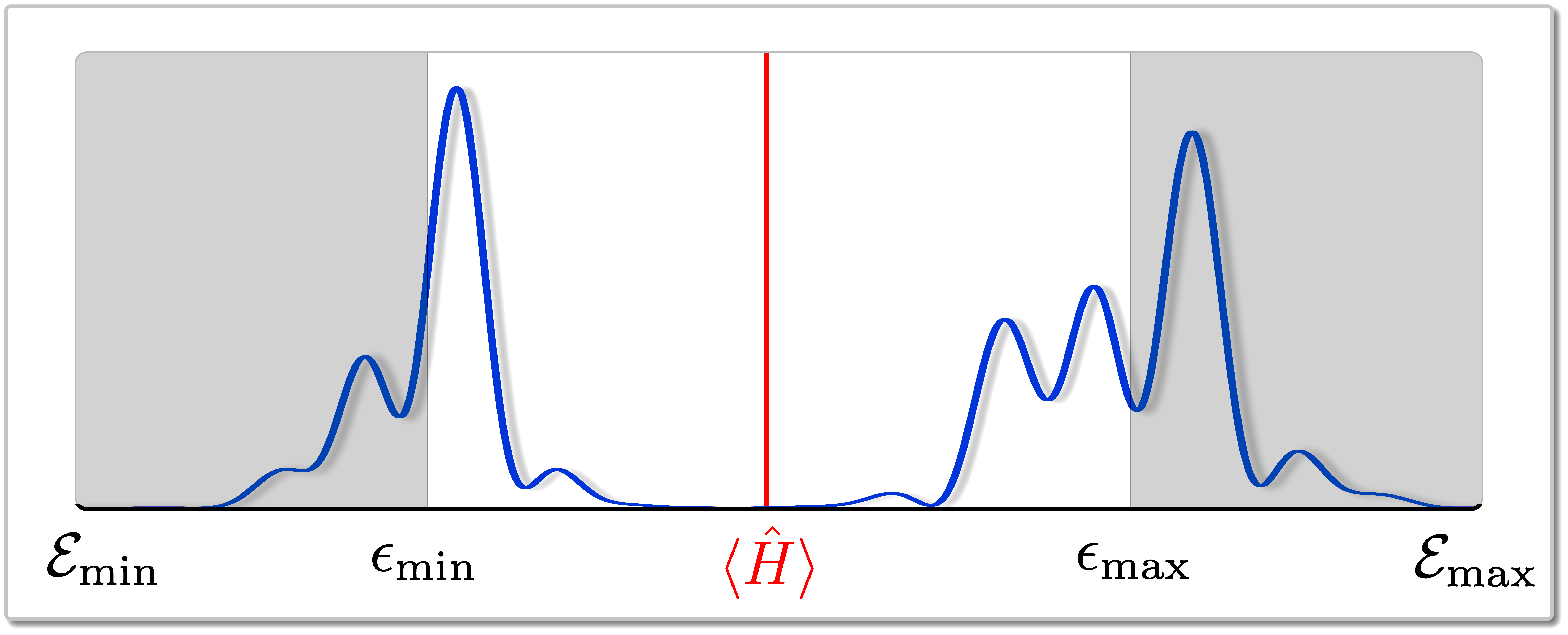}
\caption{The energy-eigenstate decomposition (blue) of a state in the ETA-Variant I subset, has an average energy (red) that falls within the energy-window (white).}
\label{fig:ewindVarI} \vf

\includegraphics[scale=0.040]{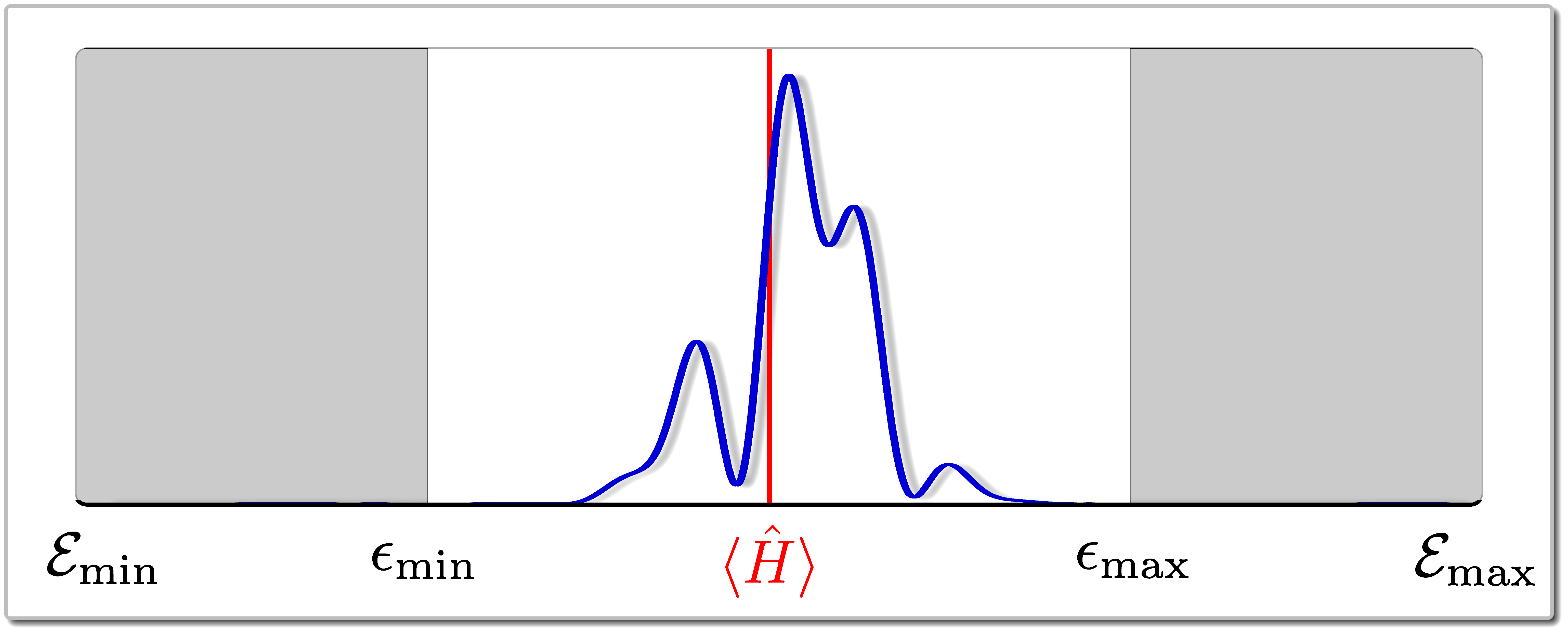}
\caption{The energy-eigenstate decomposition (blue) of a state in the ETA-Variant II subspace, is zero outside of the energy-window (white).}
\label{fig:ewindVarII}
\end{figure}

In \vIIT, the subspace mapping uses the eigenstates of the Hamiltonian to define a subspace \tsz{$\Hm_P^{(II)}\in \Hm$}. States with eigenvalues inside the energy-window are members of the subspace (\aref{fig:ewindVarII}):

%\vEq

\begin{align}
 \h H\kt{\alp_i}= \Em_{\Scale[0.8]{\alp}_{\Scale[0.7]{i}}}& \kt{\alp_i}, \ \Em_{\Scale[0.8]{\alp}_{\Scale[0.7]{i}}}\in \Dt_{\ep} \\[0.8em]
\Hm^{(II)}_P=\overline{\tt{span}}\3\{\hsh &\kt{\alp_1},\hsh \kt{\alp_2}, \ \Scale[1.12]{\hdots} \ \3\}
\end{align}

%\begin{align}
%\SDrop{3pt}{ \h P_{\hsnh \Dt_{\Scale[0.8]{\ep}}}}&=\h \iD-\sum_{\alp}\kt{\alp}\hsnh \b{\alp}, \ \forall \bk{\alp|\hs \h H \hs |\alp} \notin \Dt_{\ep} \\
%& \b{\9}\SDrop{3pt}{ \h P_{\hsnh \Dt_{\Scale[0.8]{\ep}}}}\kt{\9}=1, \ \forall \kt{\9} \in \Hm^{(II)}_P
%\end{align}
%
%\begin{equation}
%\Hm^{(II)}_P=\overline{\tt{span}}\3\{\hsh \kt{\alp_1},\hsh \hdots, \hsh \kt{\alp_N}\hsh \3\}, \ \forall \bk{\alp_i|\hs \h H \hs |\alp_i} \in \Dt_{\ep}
%\end{equation}

\subsubsection{Energy-Window Selection}

To select an energy-window, the \ti{extremum energy eigenvalues} must be determined. The \ti{extremizing orthodox mixed state} is the following:

\vEq

\begin{equation}
\td \rho_{\tt{ext}}^{(\tau)}=\fr{1}{\Zm}\hs \SRaise{4pt}{e^{\hshh \Scale[0.92]{\h H\tau}}}
\end{equation}

Taking \tsz{$\tau\ra-\infty$} results in the \ti{minimum energy}:

\begin{equation}
\lim_{\tau\ra -\infty} \tr\4[\rho_{\tt{ext}}^{(\tau)} \hs \h H\4]=\Em_{\tt{min}}	
\end{equation}

\vt

Taking \tsz{$\tau\ra \infty$} results in the \ti{maximum energy}:

\begin{equation}
\lim_{\tau\ra\infty} \tr\4[\td\rho_{\tt{ext}}^{(\tau)} \hs \h H\4]=\Em_{\tt{max}}	
\end{equation}

Viable energy-windows have the following restriction:

\vEq

\begin{equation}
\Em_{\tt{min}}<\ep_{\tt{min}}<\ep_{\tt{max}}<\Em_{\tt{max}}	
\end{equation}

\subsection{Standardization}

In \vITO and \vIIT, benchmark orthodox mixed states are generated to establish a standard of comparison.

\vFo

%
%\begin{equation}
%\ep_{\tt{min}}> \Em_{\tt{min}}, \ \ep_{\tt{max}}<\Em_{\tt{max}}
%\end{equation}

\subsubsection{Variant I}

In \vIT, orthodox mixed states are generated pseudo-randomly using positive-real functions \tsz{$f_{\gm}(x)$}.

\vo

These functions are normalized over the extremum eigenvalues:

\vEq

\begin{equation}
\int_{\Em_{\tt{min}}}^{\Em_{\tt{max}}} d\Em \ f_{\gm}(\Em)=1
\end{equation}

The average energy of the functions lies within the energy window:

\vEq 

\begin{equation}
\int_{\Em_{\tt{min}}}^{\Em_{\tt{max}}} d\Em \ f_{\gm}(\Em)\hs \Em \hs \in \hs \Dt_{\ep}
\end{equation}

An orthodox mixed state is constructed from a chosen function \tsz{$f_{\Scale[0.8]{\gm'}}(x)$}. If \tsz{$\bk{\h H}$} falls within the energy-window, a benchmark orthodox mixed state was generated:

\vEq

\begin{align}
& \hsn \hsnh \bk{\h H} \in \Dt_{\ep}, \notag \\[0.2 em]
 \td \rho(f_{\Scale[0.8]{\gm'}}) \hs &\in  \hs \4\{\Hm\tens\SRaise{6pt}{\Hm^*}\4\}^{(I)}_P		\raisetag{0 em}
\end{align}

\subsubsection{Variant II}

In \vIIT, gaussian orthodox mixed states are generated pseudo-randomly:

\vEq

\begin{align}
\td \rho_g(\ep,\sg)&=\fr{1}{\Zm}\sum_{\alp} \Scale[1.1]{e}^{-\4(\Scale[1.3]{\fr{\Em_{\alp}-\ep}{\sg}}\4)^2} \kt{\alp}\hsnh \b{\alp}	
\end{align}

To increase the likelihood that a benchmark orthodox mixed state is generated, \tsz{$\ep$} and \tsz{$\sg$} are constrained:

\vEq

\begin{align}
\ep \in \Dt_{\ep} \ \ \hs & \\[0.4 em]
e^{-\4(\Scale[1.2]{\fr{\ep_{\tt{min}}-\ep}{\sg}}\4)^2}&\ll 1 \\[0.4em]
e^{-\4(\Scale[1.2]{\fr{\ep_{\tt{max}}-\ep}{\sg}}\4)^2}&\ll 1
\end{align}

A gaussian orthodox mixed state is constructed from the chosen parameters \tsz{$(\ep,\sg)$}. The expectation value of the Hamiltonian and its variance are determined:

\vEq

\begin{equation}
\var(\h H)=\bk{\SRaise{7pt}{\h H^{\hshh 2}}}-\SRaise{7pt}{\bk{\h H}^{\hsnh 2}}
\end{equation}

If \tsz{$\bk{\h H}$} and its uncertainty fall outside the energy-window, a benchmark orthodox mixed state was not generated:

\vEq

\begin{align}
&\bk{\h H}\pm \sqrt{\var(\h H)} \hs \notin \hs \Dt_{\ep}, \notag \\[0.4em]
& \hsh \td \rho_g(\ep, \hs \sg) \hs \notin  \hs \4\{\Hm\tens\SRaise{6pt}{\Hm^*}\4\}^{(II)}_P	\raisetag{4.5 em}
\end{align}

\subsection{Arbitration}

In \vITO and \vIIT, coarse-grained benchmarking states are prepared to establish a simulation diagnostic.

\subsubsection{State Preparation}

{\vt \centering \paragraph{Variant I} ~\\ \vfo}

In \vIT, coarse-grained benchmarking states are prepared using a \ti{variational hybrid quantum-classical algorithm} \cite{VarAlg0}. These algorithms seek to use classical optimization to prepare pure states that minimize a \ti{cost-function} \cite{VarAlg1}.

\vo

A set of pure states is identified by applying a class of unitary transformations to an \ti{edge state} \tsz{$\kt{\9_e}$}:

\vEq

\begin{equation}
\K{\9_e^{(\V p)}}=\SRaise{6pt}{\h U^{(\V p)}} \K{\9_e}	
\end{equation}

\vt

The cost-function is the absolute difference between the expectation value of the simulation Hamiltonian and the center of the energy-window:

\vEq

\begin{equation}
\Cm_e(\V p)=\5|\B{\9_e^{(\V p)}} \h H_s\K{\9_e^{(\V p)}}-\4(\fr{\ep_{\tt{max}}+\ep_{\tt{min}}}{2}\4)\5|
\end{equation}

An edge state is generated pseudo-randomly, and the variational hybrid quantum-classical algorithm is allowed to run to completion. During the course of the algorithm, coarse-grained benchmarking states may be generated.

\vo \vh

The accessible coarse-grained benchmarking states are identified by the parameters \tsz{$(e,\V p_b$)} for which the following condition holds:

\vEq

\begin{equation}
\B{\9_e^{(\V p_b)}}\h H_s \K{\9_e^{(\V p_b)}}\in \Dt_{\ep}, \ \forall \hsh \K{\9_e^{(\V p_b)}}\in \Hm^{(I)}_{\SDrop{0pt}{P_{\Scale[1]{s}}}}
\end{equation}

\vth

{\vt \centering \paragraph{Variant II} ~\\ \vfo}

In \vIIT, coarse-grained benchmarking states are prepared using an \ti{adiabatic quantum simulation algorithm} \cite{AdiabSim2}. These algorithms seek to prepare eigenstates of a Hamiltonian \cite{AdiabSim0,AdiabSim1}.

\vt

An \ti{initializing Hamiltonian} \tsz{$\h H_i$} has \ti{known eigenstates} \tsz{$\kt{\SDrop{3.8pt}{\8^{(i)}_n}}$}:

\vEq

\begin{equation}
\h H_i\K{\SDrop{3.8pt}{\8^{(i)}_n}}=\SRaise{4.8pt}{\lo_n^{(i)}} \K{\SDrop{3.8pt}{\8^{(i)}_n}}
\end{equation}

\vt

The \ti{annealing time} \tsz{$\tau$} parametrizes the \ti{preparation Hamiltonian}:

\vEq

\begin{align}
\h H^{\Scale[0.8]{(i,\tau)}}_p \hsnhh \Scale[0.9]{(t)}=\4\{\h H_s-\h H_i\4\}\3(t/\tau\3)+\h H_i
\end{align}

\vt

A set of pure states is identified by time-evolving a known eigenstate under preparation Hamiltonians:

\vTo

\begin{align}
& \hsn \qquad \hs \K{\SDrop{3.8pt}{\9^{(\Scale[0.8]{i,\tau})}_n}}	=\SRaise{6pt}{\h U^{\Scale[0.8]{(i,\tau)}}} \K{\SDrop{3.8pt}{\8^{(i)}_n}} \\[0.7em]
&\raisebox{-0.4em}{\SRaise{6pt}{\h U^{\Scale[0.8]{(i,\tau)}}}}\raisebox{-0.3em}{=} \	\raisebox{-0.4em}{\Scale[1.03]{\Tm}} \hs \Scale[0.9]{\5(} \hs \raisebox{-0.5em}{\Scale[1.1]{e}}^{\hsnh -i \Scale[1.0]{\Scale[1.2]{\int}_{\hsn 0}^{\tau} dt^{'} \ \h H^{\Scale[0.8]{(i,\tau)}}_p \hsnh \Scale[0.9]{(t^{'})}}}\hs \Scale[0.9]{\5)}	
\end{align}

\vt

The \ti{simulation eigenvalues} \tsz{$E_n^{(\Scale[0.8]{i,\tau})}$} are defined as follows:

\vEq

\begin{align}
&\qquad \ \h H_s\K{\SRaise{4.3pt}{\SDrop{3pt}{\alp^s_k}}}= \SRaise{5pt}{\Em^s_{\Scale[0.8]{\alp}_{\Scale[0.7]{k}}}} \K{\SRaise{4.3pt}{\SDrop{3pt}{\alp^s_k}}} \\[0.4em]
&\BK{\SRaise{4.3pt}{\SDrop{3pt}{\alp^s_k}} \hsh | \hsh\SDrop{3.8pt}{\9^{(\Scale[0.8]{i,\tau})}_n}}\neq 0, \ \ \forall \hsh \SRaise{5pt}{\Em^s_{\Scale[0.8]{\alp}_{\Scale[0.7]{k}}}} \in E_n^{(\Scale[0.8]{i,\tau})}
\end{align}

\vt

After specifying an initializing Hamiltonian, both the annealing time and the known eigenstate are chosen pseudo-randomly.

The accessible coarse-grained benchmarking states are identified by the parameters \tsz{$(i,\tau_b,n_b)$} for which the following condition holds:

\vEq

\begin{align}
E_{n_b}^{(\Scale[0.8]{i,\tau_b})} \in \Dt_{\ep}, \ \forall \K{\SDrop{3.8pt}{\9^{(\Scale[0.8]{i,\tau_b})}_{n_b}}} \in \Hm^{(II)}_{\SDrop{0pt}{P_{\Scale[1]{s}}}}
\end{align}

\vt

To determine if a coarse-grained benchmarking state was generated, quantum simulation is used to place a bound on the simulation eigenvalues of \tsz{$\kt{\SDrop{3.8pt}{\9^{(\Scale[0.8]{i,\tau})}_n}}$}.

\vt

The \ti{simulated expectation value} is the following:

\vEq

\begin{align}
\hsn \bk{\h O_s(t)}&=\B{\SDrop{3.8pt}{\9^{(\Scale[0.8]{i,\tau})}_n}}\SRaise{4pt}{e^{\Scale[0.92]{\hsh i\h H_s t}}}\hs \h O \hs \SRaise{4pt}{e^{-\Scale[0.92]{i\h H_s t}}}\K{\SDrop{3.8pt}{\9^{(\Scale[0.8]{i,\tau})}_n}}\\[0.4em]
&=\sum_{\SRaise{4.3pt}{\alp^s}\hsnh ,\hsh \SRaise{5pt}{\bt^s}} \SDrop{3pt}{c^*_{\SRaise{4.8pt}{\alp^s}}} \hsh \SDrop{2.8pt}{c_{ \SRaise{4.8pt}{\bt^s}}} \b{\SRaise{4.8pt}{\alp^{\hsnhh s}}}\h O \kt{\SRaise{4.3pt}{\bt^s}} \ \SRaise{4pt}{e^{-\Scale[0.92]{it\3(\Em^s_{\bt}-\Em^s_{\alp}\3)}}} \\[0.4em]
&=\sum_{\SRaise{4.3pt}{\alp^s} \hsnh ,\hsh \SRaise{5pt}{\bt^s}} \SDrop{2.5pt}{\Dm_{\hsnh \SRaise{4.8pt}{\alp^s} \hsnh , \hsh \SRaise{4.3pt}{\bt^s}}} \ \SRaise{4pt}{e^{-\Scale[0.92]{it\3(\Em^s_{\bt}-\Em^s_{\alp}\3)}}}
\end{align}

The Fourier transform of the simulated expectation value is the following:

\vEq

\begin{align}
\bk{\h O_s(\om)}&=\int_{-\infty}^{\infty} dt \ \SRaise{4pt}{e^{-\Scale[0.85]{i\om t}}} \bk{\h O_s(t)} \\[0.4em]
&=\sum_{\SRaise{4.3pt}{\alp^s} \hsnh ,\hsh \SRaise{5pt}{\bt^s}} \SDrop{2.5pt}{\Dm_{\hsnh \SRaise{4.8pt}{\alp^s} \hsnh , \hsh \SRaise{4.3pt}{\bt^s}}} \ \dt\4\{\om-\4(\Em^s_{\alp}-\Em^s_{\bt}\4)\4\} \\[0.4em]
&=\sum_{\SRaise{4.3pt}{\alp^s} \hsnh ,\hsh \SRaise{5pt}{\bt^s}} \SDrop{2.5pt}{\Dm_{\hsnh \SRaise{4.8pt}{\alp^s} \hsnh , \hsh \SRaise{4.3pt}{\bt^s}}} \ \dt\4\{\om-\Dt^{\hsnh s}_{\SRaise{4.8pt}{\alp^s} \hsnh , \hsh \SRaise{4.3pt}{\bt^s}}\4\}
\end{align}

The Fourier transform is peaked at the \ti{energy gaps} (\tsz{$\Dt^{\hsnh s}_{\SRaise{4.8pt}{\alp^s} \hsnh , \hsh \SRaise{4.3pt}{\bt^s}}$}) of the simulation eigenvalues. The peaks are discernible if \tsz{$\b{\SRaise{4.8pt}{\alp^{\hsnhh s}}}\h O \kt{\SRaise{4.3pt}{\bt^s}}\neq 0$}.

\vt

The \ti{maximum energy-gap} \tsz{$\Dt^{\hsnh s}_{\tt{max}}$} can be used to place a bound on the simulation eigenvalues:

\vEq

\begin{align}
\bk{\h H_s}-\Dt^{\hsnh s}_{\tt{max}}\leq \ E_n^{(\Scale[0.8]{i,\tau})} \leq \bk{\h H_s}+\Dt^{\hsnh s}_{\tt{max}}
\end{align}

\vt

If the simulation eigenvalue bound falls within the energy-window, a coarse-grained benchmarking state was prepared:

\vEq

\begin{equation}
\bk{\h H_s}\pm \Dt^{\hsnh s}_{\tt{max}} \in \Dt_{\ep}, \ \hs \K{\SDrop{3.8pt}{\9^{(i,\tau)}_n}} \in	\Hm^{(II)}_{\SDrop{0pt}{P_{\Scale[1]{s}}}}
\end{equation}

\subsubsection{Time-Averaging}

In \vITO and \vIIT, the simulated time-averaged expectation values of the coarse-grained benchmarking states must be estimated.

\vo

The simulated expectation value averaged over time \tsz{$t_a$} is the following:

\vEq

\begin{align}
\bk{\h O_s}_{\tt{ave}}&=\fr{1}{t_a} \int_0^{t_a}dt^{'} \ \bk{\h O_s(t^{'})} \notag \\
&=\sum_{\SRaise{4.3pt}{\alp^s} \hsnh ,\hsh \SRaise{5pt}{\bt^s}} \SDrop{2.5pt}{\Dm_{\hsnh \SRaise{4.8pt}{\alp^s} \hsnh , \hsh \SRaise{4.3pt}{\bt^s}}} \ \fr{1}{t_a}\int_0^{t_a} dt^{'} \ \SRaise{4pt}{e^{\hsh \Scale[0.92]{it^{'}\3(\Dt^{\hsnh s}_{\SRaise{4.8pt}{\alp^s} \hsnh , \hsh \SRaise{4.3pt}{\bt^s}}\3)}}}
\end{align}

\vt

The simulated time-averaged expectation value is well-approximated when \tsz{$t_a$} is much larger than the \ti{maximum period}, which is set by the \ti{minimum energy-gap} \tsz{$\Dt^{\hsnh s}_{\tt{min}}$}:\footnote{This scaling may be considerably relaxed when the \ti{Eigenstate Thermalization Hypothesis} (ETH) is valid \cite{eth0,eth1,eth2,eth3,eth4,eth4a,eth5,eth6,eth7,eth8,eth9,eth10,eth11,eth12,eth13,eth14,eth15,eth16,eth17,eth18,eth19,eth20}.}

\vEq

\begin{align}
& t_a \gg \fr{2\pi}{\Dt^{\hsnh s}_{\tt{min}}}>\fr{2\pi}{\Dt^{\hsnh s}_{\tt{max}}}, \notag \\[0.6em]
&  \ \ \  \bk{\h O_s}_{\tt{ave}}\approx\bk{\h \Om_s}
\end{align}

\vo

To resolve the dynamics, the simulated expectation value must be sampled faster than the \ti{aliasing time}, which is set by the maximum energy-gap \cite{Nyquist0,Nyquist1,Nyquist2}:

\vEq

\begin{equation}
t_{samp}<\fr{\pi}{\Dt^{\hsnh s}_{\tt{max}}}	
\end{equation}

\vt

When \tsz{$\h H_s\ra\h H$}, the maximum energy-gap is upper-bounded by the \ti{energy breadth}:

\vEq

\begin{equation}
\Dt_{B}=\Em_{\tt{max}}-\Em_{\tt{min}}	
\end{equation}

\vt
%
%The extremum energy-gap is used to set the following constraints on the simulation parameters:

%The simulation parameters are constrained by the extremum energy-gap:

The energy breadth is used to place constraints on the time-averaging parameters:

\vEq

\begin{align}
t_{\tt{samp}}< \fr{\pi}{\Dt_{B}} \\[0.7em]
t_a\gg \fr{2\pi}{\Dt_{B}}
\end{align}

% \tsz{$t_{\tt{samp}}< \pi/\Dt^{\hsnh \Scale[0.6]{(H)}}_{\tt{max}}$}. The time-average is taken over \tsz{$t_a\gg 2\pi/\Dt^{\hsnh \Scale[0.6]{(H)}}_{\tt{max}}$}.

\section{Numerical Implementation}

To examine their efficacy, both \vITO and \vIITO are applied to a \ti{Hamiltonian family}.

\subsection{Benchmark Procedure}

\subsubsection{Hamiltonian Family}

The Hamiltonian family describes a particle of mass \tsz{$m$}, with a kinetic term that has a length-scale \tsz{$a$}:

\vEq

\begin{equation}
\h H_f(\om,\lo)=-\fr{1}{2m} \5[\fr{\SRaise{4pt}{e^{-\Scale[0.92]{i \h p a}}}-\SRaise{4pt}{e^{\Scale[0.92]{i \h p a}}}}{2a}\5]^2+\fr{1}{2}\om \hs \SRaise{4pt}{\h x^2} +\lo \hs \SRaise{4pt}{\h x^3}
\end{equation}

\vt

The particle is confined to sites in a periodic lattice:

\vTh

\begin{equation}
\tt{sites}:\{-aN,\hsh a\hshh (-N+1\hshh) \hshh,\ \Scale[1.12]{\hdots} \ , \hsh a\hshh (N-1 \hshh) \hshh, \hsh aN\}	
\end{equation}

\subsubsection{Benchmark Hamiltonians}

\benchMTO is used to distinguish members of the Hamiltonian family from one another. $1001$-site lattices are used. 

\vt

The \ti{target Hamiltonian} has a fully quadratic potential:

\vTh

\begin{equation}
\h H=\h H_f(\om_0,0)	
\end{equation}

\vt

The \ti{corrupted Hamiltonians} are generated by adjusting the cubic potential:

\vTh

\begin{equation}
\h H_c(\lo)= \h H_f(\om_0,\lo)	
\end{equation}

\vt

The \ti{corruption strength} is the following quantity:

\vTh

\begin{equation}
\eta=a\fr{\lo}{\om_0}	
\end{equation}

\subsubsection{Energy-Window Selection}

The \ti{target energy-range} \tsz{$\Dt_{\Scale[0.8]{\Em}}$} is defined as follows:

\vTh

\begin{equation}
\Em_{\tt{min}}\leq \Em \leq \Em_{\tt{max}}, \ \forall \Em \in \Dt_{\Scale[0.8]{\Em}}
\end{equation}

\vt

Three energy-windows are used during the benchmark:

\vspace{-5mm}

\begin{align*}
\tt{\ti{Low-Range}: } \tt{bottom-15\% of the target energy-range} \\
\tt{\ti{Mid-Range}: } \tt{median-15\% of the target energy-range} \\	
\tt{\ti{High-Range}: } \tt{top-15\% of the target energy-range}
\end{align*}

\subsubsection{Observable Selection}

The benchmark observable is the following:

\vEq

\begin{equation}
\h X_{a}=\SRaise{6pt}{\3|\hsh \h x\hsh \3|^{1/3}}	
\end{equation}

\subsubsection{Standardization}

%The target energy-range is determined by sampling \tsz{$\SRaise{4pt}{10^7}$} configurations from the extremizing state.

%\vt

To establish a standard of comparison, \tsz{$\SRaise{4pt}{10^4}$} benchmark orthodox mixed states are generated. A \ti{bootstrapping algorithm} is used for statistical analysis \cite{bootstrapI,bootstrapII,bootstrapIII}.

\subsubsection{Arbitration}

To establish a simulation diagnostic, \tsz{$\SRaise{4pt}{10^4}$} coarse-grained benchmarking states are generated. A bootstrapping algorithm is used for statistical analysis.

\subsection{Numerical Results}

\subsubsection{Low-Range Window}

\tus{Variant I}: \vITO positively benchmarks the target Hamiltonian. \vITO  negatively benchmarks \tsz{$3$} out of \tsz{$5$} corrupted Hamiltonians (\aref{fig:varoneR0}).

\vt

\tus{Variant II}: \vIITO positively benchmarks the target Hamiltonian. \vIITO negatively benchmarks \tsz{$5$} out of \tsz{$5$} corrupted Hamiltonians (\aref{fig:vartwoR0}).

\subsubsection{Mid-Range Window}

\tus{Variant I}: \vITO positively benchmarks the target Hamiltonian. \vITO negatively benchmarks \tsz{$5$} out of \tsz{$5$} corrupted Hamiltonians (\aref{fig:varoneR1}).

\vt

\tus{Variant II}: \vIITO positively benchmarks the target Hamiltonian. \vIITO negatively benchmarks \tsz{$5$} out of \tsz{$5$} corrupted Hamiltonians (\aref{fig:vartwoR1}).

\subsubsection{High-Range Window}

\tus{Variant I}: \vITO positively benchmarks the target Hamiltonian. \vITO negatively benchmarks \tsz{$5$} out of \tsz{$5$} corrupted Hamiltonians (\aref{fig:varoneR2}).

\vt

\tus{Variant II}: \vIITO positively benchmarks the target Hamiltonian. \vIITO negatively benchmarks \tsz{$5$} out of \tsz{$5$} corrupted Hamiltonians (\aref{fig:vartwoR2}).

%\appendix\label{appendix:ref}
\renewcommand\thesubsection{\Roman{subsection}}

%\begin{appendices}

\section{Appendix: Monte-Carlo Methods}\label{appendix:AMonteCarlo}

%\vTh

\ti{Monte-Carlo techniques} are a class of algorithms that employ successive random sampling \cite{MonteTech0}. In particular,  \ti{Markov-chain Monte-Carlo methods} approximate sampling from a target probability distribution by recording the output of a stochastic numerical simulation \cite{Metro0,Metro1,Metro2,Metro3}. 

\vt

As the simulation time \tsz{$\tau_s$} tends to infinity, the simulated distribution recovers the target distribution. The rate of convergence is independent of dimension: \tsz{$\sim \Oe(1/\sqrt{\tau_s})$} \cite{MonteCarlo0}. The classical computations required for this work are \ti{tractable} \cite{cobhamSim}, even if quantum simulation is unfeasible due to the Hilbert space dimension \cite{feynmansupremacySim,complexTwo,completeIV,completeI,completeII,interI}.

%\end{appendices}

\section{Acknowledgements}\label{sec:ack}

\vTh

\begin{center}

\noindent\rule{3cm}{0.4pt}

\end{center}

\begin{center}
\ti{All of us, like sheep, have strayed away;  \newline
we have left God's paths to follow our own. \newline
Yet the Lord laid on Him the sins of us all.}

\vo

-\ti{Isaiah 53:6}
\end{center}

\begin{center}
$ -AMDG - $
\end{center}

\clearpage

\setcounter{figure}{4}

\begin{figure}[H]
\includegraphics[scale=0.215]{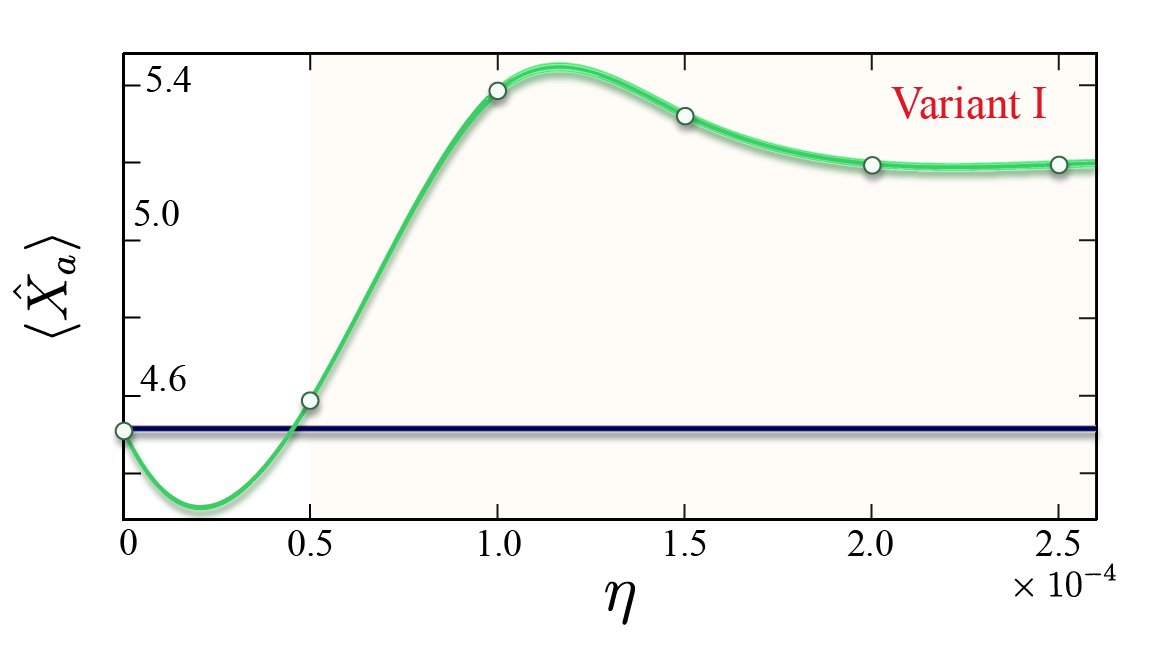}
\caption{ \ti{Low-Range}: The standard of comparison for the target Hamiltonian is determined using classical computing resources (blue curve). It is contrasted against the simulation diagnostic for the corrupted Hamiltonians (green curve).}
\label{fig:varoneR0}
\end{figure}

\setcounter{figure}{6}

\begin{figure}[H]
\includegraphics[scale=0.215]{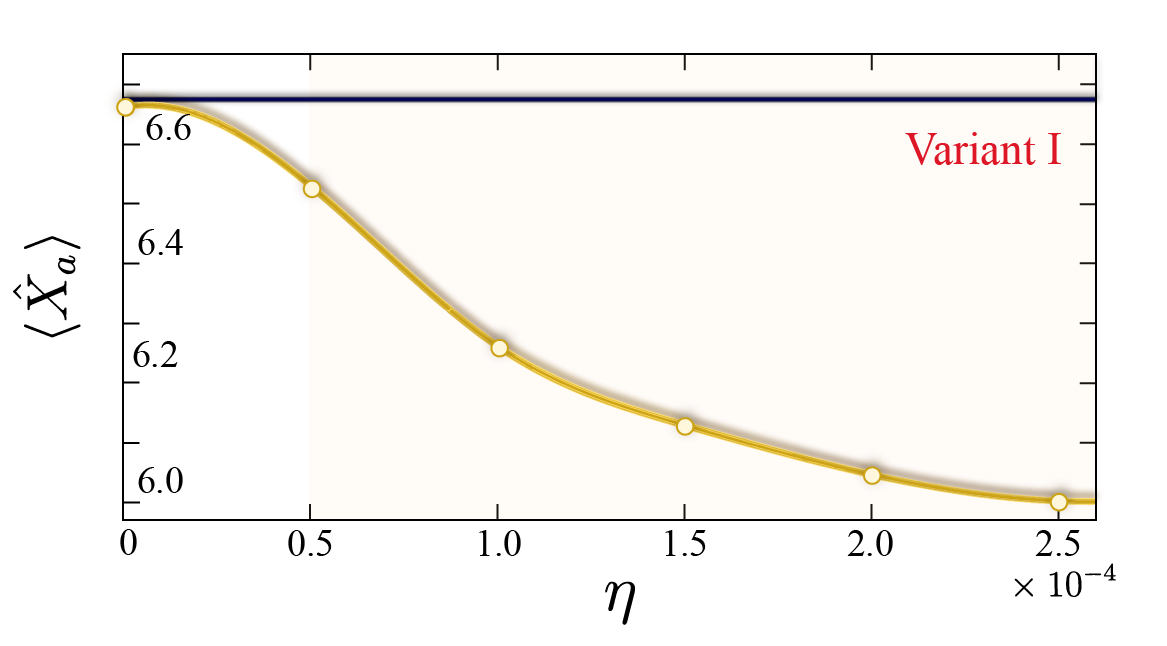}
\caption{ \ti{Mid-Range}: \hsh The standard of comparison for the target Hamiltonian is determined using classical computing resources (blue curve). It is contrasted against the simulation diagnostic for the corrupted Hamiltonians (yellow curve).}
\label{fig:varoneR1}
\end{figure}

\setcounter{figure}{8}
 
 \begin{figure}[H]
\includegraphics[scale=0.215]{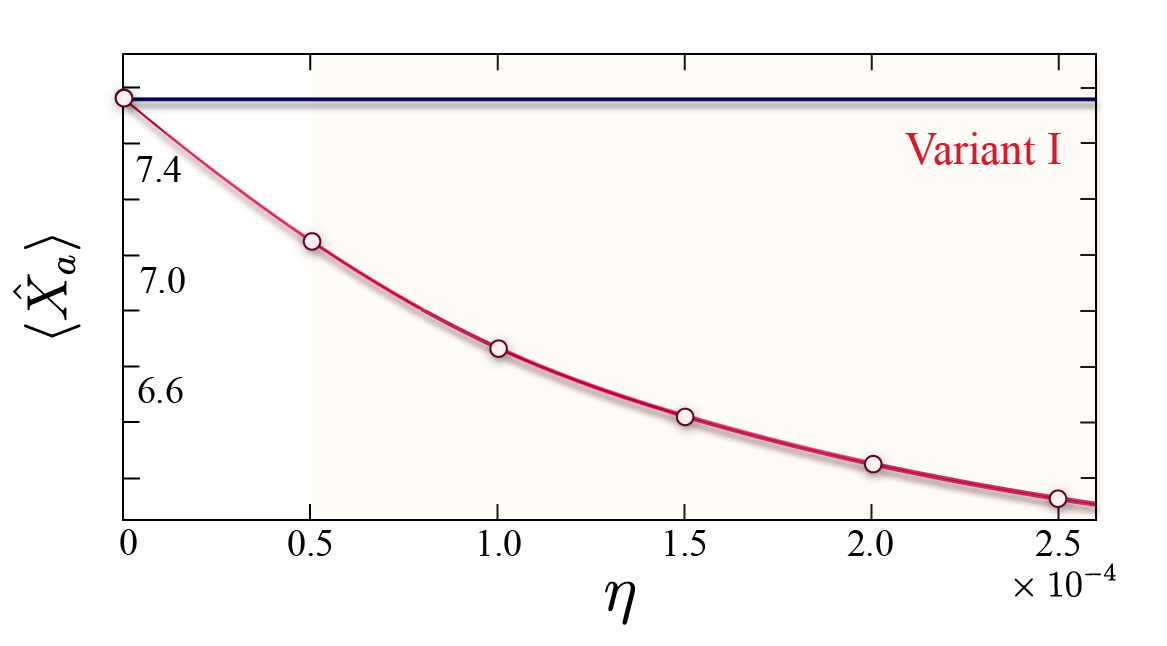}
\caption{ \ti{High-Range}:  The standard of comparison for the target Hamiltonian is determined using classical computing resources (blue curve). It is contrasted against the simulation diagnostic for the corrupted Hamiltonians (red curve).}
\label{fig:varoneR2}
\end{figure}

\newpage

\setcounter{figure}{5}

\begin{figure}[H]
\includegraphics[scale=0.215]{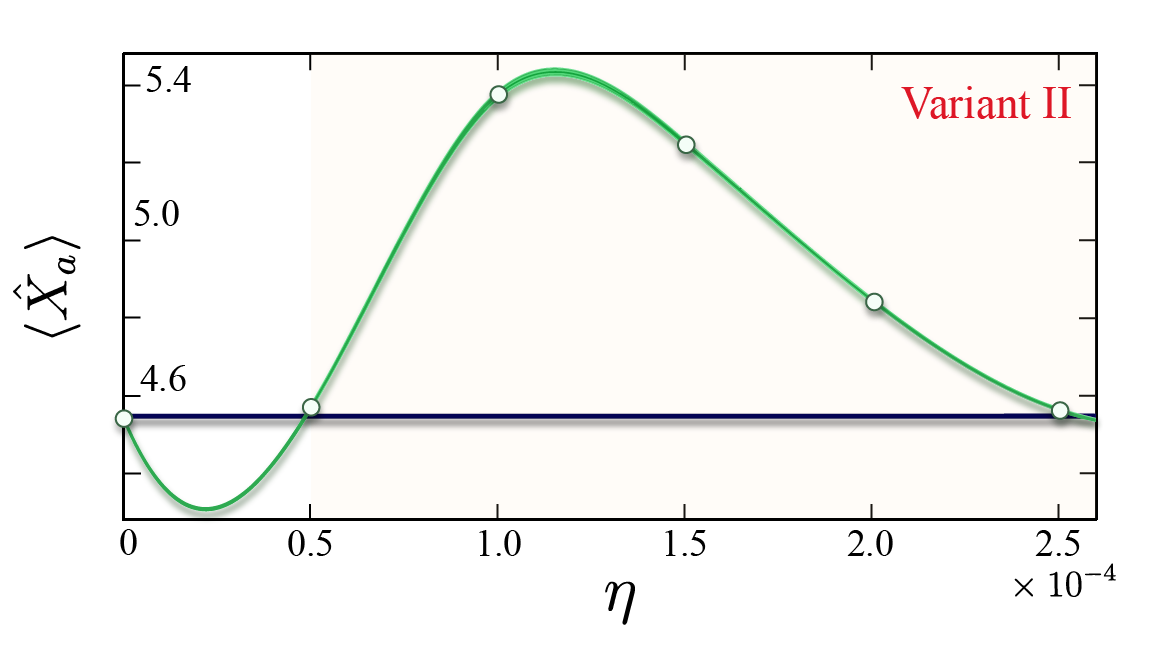}
\caption{ \ti{Low-Range}: The standard of comparison for the target Hamiltonian is determined using classical computing resources (blue curve). It is contrasted against the simulation diagnostic for the corrupted Hamiltonians (green curve).}
\label{fig:vartwoR0}
\end{figure}

\setcounter{figure}{7}

\begin{figure}[H]
\includegraphics[scale=0.215]{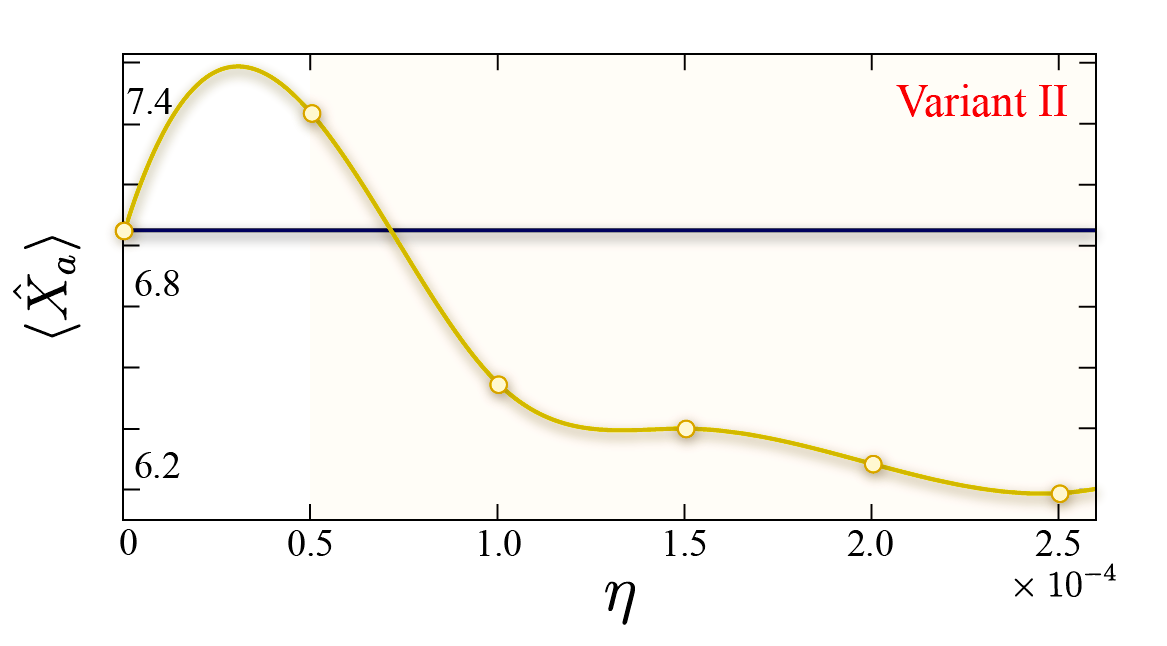}
\caption{ \ti{Mid-Range}: \hsh The standard of comparison for the target Hamiltonian is determined using classical computing resources (blue curve). It is contrasted against the simulation diagnostic for the corrupted Hamiltonians (yellow curve).}
\label{fig:vartwoR1}
\end{figure}

\setcounter{figure}{9}

\begin{figure}[H]
\includegraphics[scale=0.215]{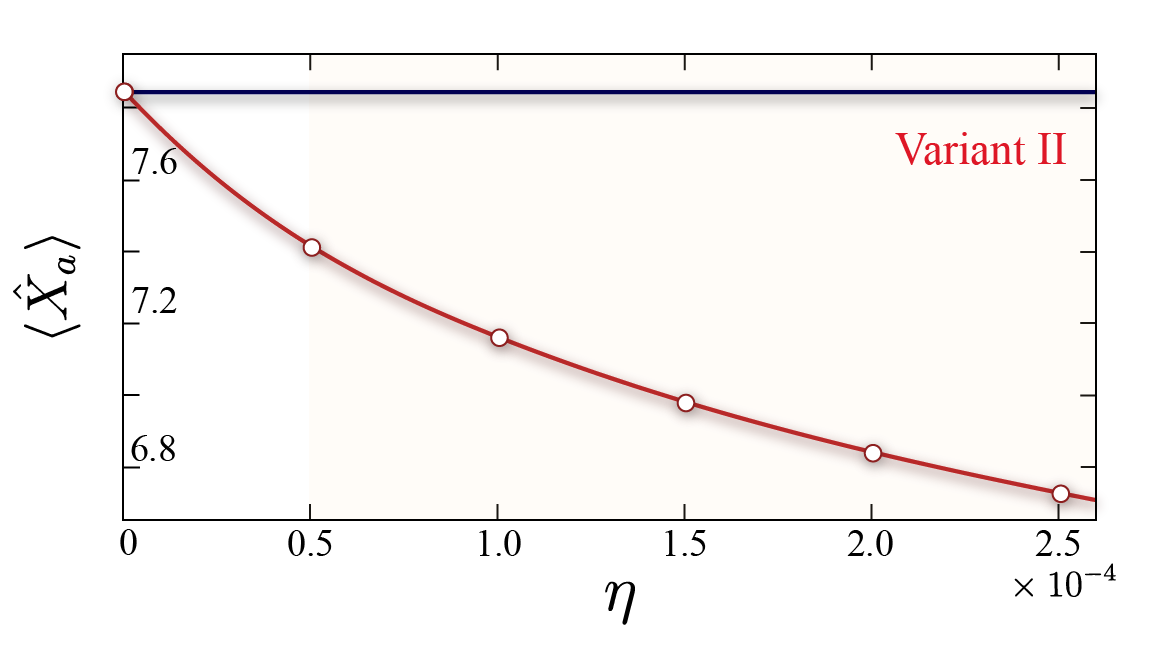}
\caption{ \ti{High-Range}:  The standard of comparison for the target Hamiltonian is determined using classical computing resources (blue curve). It is contrasted against the simulation diagnostic for the corrupted Hamiltonians (red curve).}
\label{fig:vartwoR2}
\end{figure}

\rhead{}

\clearpage

\rhead{REFERENCES}

\renewcommand*{\bibfont}{\scriptsize}

\bibliography{ETHref}

\providecommand{\noopsort}[1]{}\providecommand{\singleletter}[1]{#1}%
\begin{thebibliography}{58}%
\makeatletter
\providecommand \@ifxundefined [1]{%
 \@ifx{#1\undefined}
}%
\providecommand \@ifnum [1]{%
 \ifnum #1\expandafter \@firstoftwo
 \else \expandafter \@secondoftwo
 \fi
}%
\providecommand \@ifx [1]{%
 \ifx #1\expandafter \@firstoftwo
 \else \expandafter \@secondoftwo
 \fi
}%
\providecommand \natexlab [1]{#1}%
\providecommand \enquote  [1]{``#1''}%
\providecommand \bibnamefont  [1]{#1}%
\providecommand \bibfnamefont [1]{#1}%
\providecommand \citenamefont [1]{#1}%
\providecommand \href@noop [0]{\@secondoftwo}%
\providecommand \href [0]{\begingroup \@sanitize@url \@href}%
\providecommand \@href[1]{\@@startlink{#1}\@@href}%
\providecommand \@@href[1]{\endgroup#1\@@endlink}%
\providecommand \@sanitize@url [0]{\catcode `\\12\catcode `\$12\catcode
  `\&12\catcode `\#12\catcode `\^12\catcode `\_12\catcode `\%12\relax}%
\providecommand \@@startlink[1]{}%
\providecommand \@@endlink[0]{}%
\providecommand \url  [0]{\begingroup\@sanitize@url \@url }%
\providecommand \@url [1]{\endgroup\@href {#1}{\urlprefix }}%
\providecommand \urlprefix  [0]{URL }%
\providecommand \Eprint [0]{\href }%
\providecommand \doibase [0]{https://doi.org/}%
\providecommand \selectlanguage [0]{\@gobble}%
\providecommand \bibinfo  [0]{\@secondoftwo}%
\providecommand \bibfield  [0]{\@secondoftwo}%
\providecommand \translation [1]{[#1]}%
\providecommand \BibitemOpen [0]{}%
\providecommand \bibitemStop [0]{}%
\providecommand \bibitemNoStop [0]{.\EOS\space}%
\providecommand \EOS [0]{\spacefactor3000\relax}%
\providecommand \BibitemShut  [1]{\csname bibitem#1\endcsname}%
\let\auto@bib@innerbib\@empty
%</preamble>
\bibitem [{\citenamefont {Schr{\"o}dinger}(1926)}]{tame0}%
  \BibitemOpen
  \bibfield  {author} {\bibinfo {author} {\bibfnamefont {E.}~\bibnamefont
  {Schr{\"o}dinger}},\ }\bibfield  {title} {\bibinfo {title} {Quantization as
  an eigenvalue problem},\ }\href
  {https://doi.org/https://doi.org/10.1002/andp.19263840404} {\bibfield
  {journal} {\bibinfo  {journal} {Annalen der Physik}\ }\textbf {\bibinfo
  {volume} {384}},\ \bibinfo {pages} {361} (\bibinfo {year}
  {1926})}\BibitemShut {NoStop}%
\bibitem [{\citenamefont {Neumann}(1927)}]{densmat0}%
  \BibitemOpen
  \bibfield  {author} {\bibinfo {author} {\bibfnamefont {J.~v.}\ \bibnamefont
  {Neumann}},\ }\bibfield  {title} {\bibinfo {title} {Probabilistic theory of
  quantum mechanics},\ }\href {http://eudml.org/doc/59230} {\bibfield
  {journal} {\bibinfo  {journal} {Nachrichten von der Gesellschaft der
  Wissenschaften zu G{\"o}ttingen, Mathematisch-Physikalische Klasse}\ }\textbf
  {\bibinfo {volume} {1927}},\ \bibinfo {pages} {245} (\bibinfo {year}
  {1927})}\BibitemShut {NoStop}%
\bibitem [{\citenamefont {Landau}(1927)}]{densmat1}%
  \BibitemOpen
  \bibfield  {author} {\bibinfo {author} {\bibfnamefont {L.}~\bibnamefont
  {Landau}},\ }\bibfield  {title} {\bibinfo {title} {The damping problem in
  quantum mechanics},\ }in\ \href
  {https://doi.org/https://doi.org/10.1016/B978-0-08-010586-4.50007-9} {\emph
  {\bibinfo {booktitle} {Collected Papers of L.D. Landau}}}\ (\bibinfo
  {publisher} {Pergamon},\ \bibinfo {year} {1927})\ Chap.~\bibinfo {chapter}
  {2}, pp.\ \bibinfo {pages} {8--18}\BibitemShut {NoStop}%
\bibitem [{\citenamefont {Nielsen}\ and\ \citenamefont
  {Chuang}(2011{\natexlab{a}})}]{densmat2}%
  \BibitemOpen
  \bibfield  {author} {\bibinfo {author} {\bibfnamefont {M.~A.}\ \bibnamefont
  {Nielsen}}\ and\ \bibinfo {author} {\bibfnamefont {I.~L.}\ \bibnamefont
  {Chuang}},\ }\href@noop {} {\emph {\bibinfo {title} {Quantum computation and
  quantum information}}}\ (\bibinfo  {publisher}
  {\href{http://mmrc.amss.cas.cn/tlb/201702/W020170224608149940643.pdf}{Cambridge
  University Press}},\ \bibinfo {year} {2011})\ pp.\ \bibinfo {pages}
  {98--108}\BibitemShut {NoStop}%
\bibitem [{\citenamefont {{Fubini}}(1904)}]{haar0}%
  \BibitemOpen
  \bibfield  {author} {\bibinfo {author} {\bibfnamefont {G.}~\bibnamefont
  {{Fubini}}},\ }\bibfield  {title} {\bibinfo {title} {On the metrics defined
  by a hermitian form},\ }\href@noop {} {\bibfield  {journal} {\bibinfo
  {journal}
  {\href{https://books.google.com/books/about/Sulle_metriche_definite_da_una_forma_her.html?id=ljeNnQAACAAJ}{Office
  graf. C. Ferrari}}\ } (\bibinfo {year} {1904})}\BibitemShut {NoStop}%
\bibitem [{\citenamefont {Study}(1905)}]{haar1}%
  \BibitemOpen
  \bibfield  {author} {\bibinfo {author} {\bibfnamefont {E.}~\bibnamefont
  {Study}},\ }\bibfield  {title} {\bibinfo {title} {Shortest routes in a
  complex area},\ }\href {https://doi.org/10.1007/BF01457616} {\bibfield
  {journal} {\bibinfo  {journal} {Mathematische Annalen}\ }\textbf {\bibinfo
  {volume} {60}},\ \bibinfo {pages} {321} (\bibinfo {year} {1905})}\BibitemShut
  {NoStop}%
\bibitem [{\citenamefont {Haar}(1933)}]{haar2}%
  \BibitemOpen
  \bibfield  {author} {\bibinfo {author} {\bibfnamefont {A.}~\bibnamefont
  {Haar}},\ }\bibfield  {title} {\bibinfo {title} {The standard in the theory
  of continuous groups},\ }\href {https://doi.org/10.2307/1968346} {\bibfield
  {journal} {\bibinfo  {journal} {Annals of Mathematics}\ }\textbf {\bibinfo
  {volume} {34}},\ \bibinfo {pages} {147} (\bibinfo {year} {1933})}\BibitemShut
  {NoStop}%
\bibitem [{\citenamefont {Boya}\ \emph {et~al.}(2003)\citenamefont {Boya},
  \citenamefont {Sudarshan},\ and\ \citenamefont {Tilma}}]{haar3}%
  \BibitemOpen
  \bibfield  {author} {\bibinfo {author} {\bibfnamefont {L.~J.}\ \bibnamefont
  {Boya}}, \bibinfo {author} {\bibfnamefont {E.}~\bibnamefont {Sudarshan}},\
  and\ \bibinfo {author} {\bibfnamefont {T.}~\bibnamefont {Tilma}},\ }\bibfield
   {title} {\bibinfo {title} {Volumes of compact manifolds},\ }\href
  {https://doi.org/https://doi.org/10.1016/S0034-4877(03)80038-1} {\bibfield
  {journal} {\bibinfo  {journal} {Reports on Mathematical Physics}\ }\textbf
  {\bibinfo {volume} {52}},\ \bibinfo {pages} {401} (\bibinfo {year}
  {2003})}\BibitemShut {NoStop}%
\bibitem [{\citenamefont {Feynman}(1982)}]{feynmansupremacySim}%
  \BibitemOpen
  \bibfield  {author} {\bibinfo {author} {\bibfnamefont {R.~P.}\ \bibnamefont
  {Feynman}},\ }\bibfield  {title} {\bibinfo {title} {Simulating physics with
  computers},\ }\href {https://doi.org/10.1007/BF02650179} {\bibfield
  {journal} {\bibinfo  {journal} {International Journal of Theoretical
  Physics}\ }\textbf {\bibinfo {volume} {21}},\ \bibinfo {pages} {467}
  (\bibinfo {year} {1982})}\BibitemShut {NoStop}%
\bibitem [{\citenamefont {{Manin}}(1980)}]{maninsupremacySim}%
  \BibitemOpen
  \bibfield  {author} {\bibinfo {author} {\bibfnamefont {Y.~I.}\ \bibnamefont
  {{Manin}}},\ }\bibfield  {title} {\bibinfo {title} {Computable and
  uncomputable},\ }\href@noop {} {\bibfield  {journal} {\bibinfo  {journal}
  {\href{https://bit.ly/YuriManinSim}{Sovetskoye Radio, Moscow}}\ } (\bibinfo
  {year} {1980})}\BibitemShut {NoStop}%
\bibitem [{\citenamefont {Turing}(1937)}]{ClassComp0}%
  \BibitemOpen
  \bibfield  {author} {\bibinfo {author} {\bibfnamefont {A.~M.}\ \bibnamefont
  {Turing}},\ }\bibfield  {title} {\bibinfo {title} {On computable numbers,
  with an application to the decision problem},\ }\href
  {https://doi.org/10.1112/plms/s2-42.1.230} {\bibfield  {journal} {\bibinfo
  {journal} {Proceedings of the London Mathematical Society}\ }\textbf
  {\bibinfo {volume} {s2-42}},\ \bibinfo {pages} {230} (\bibinfo {year}
  {1937})}\BibitemShut {NoStop}%
\bibitem [{\citenamefont {Feynman}(1948)}]{pathint0}%
  \BibitemOpen
  \bibfield  {author} {\bibinfo {author} {\bibfnamefont {R.~P.}\ \bibnamefont
  {Feynman}},\ }\bibfield  {title} {\bibinfo {title} {Space-time approach to
  non-relativistic quantum mechanics},\ }\href
  {https://doi.org/10.1103/RevModPhys.20.367} {\bibfield  {journal} {\bibinfo
  {journal} {Rev. Mod. Phys.}\ }\textbf {\bibinfo {volume} {20}},\ \bibinfo
  {pages} {367} (\bibinfo {year} {1948})}\BibitemShut {NoStop}%
\bibitem [{\citenamefont {Nielsen}\ and\ \citenamefont
  {Chuang}(2011{\natexlab{b}})}]{projec0}%
  \BibitemOpen
  \bibfield  {author} {\bibinfo {author} {\bibfnamefont {M.~A.}\ \bibnamefont
  {Nielsen}}\ and\ \bibinfo {author} {\bibfnamefont {I.~L.}\ \bibnamefont
  {Chuang}},\ }\href@noop {} {\emph {\bibinfo {title} {Quantum computation and
  quantum information}}}\ (\bibinfo  {publisher}
  {\href{http://mmrc.amss.cas.cn/tlb/201702/W020170224608149940643.pdf}{Cambridge
  University Press}},\ \bibinfo {year} {2011})\ pp.\ \bibinfo {pages}
  {70--72}\BibitemShut {NoStop}%
\bibitem [{\citenamefont {Peruzzo}\ \emph {et~al.}(2014)\citenamefont
  {Peruzzo}, \citenamefont {McClean}, \citenamefont {Shadbolt}, \citenamefont
  {Yung}, \citenamefont {Zhou}, \citenamefont {Love}, \citenamefont
  {Aspuru-Guzik},\ and\ \citenamefont {O'Brien}}]{VarAlg0}%
  \BibitemOpen
  \bibfield  {author} {\bibinfo {author} {\bibfnamefont {A.}~\bibnamefont
  {Peruzzo}}, \bibinfo {author} {\bibfnamefont {J.}~\bibnamefont {McClean}},
  \bibinfo {author} {\bibfnamefont {P.}~\bibnamefont {Shadbolt}}, \bibinfo
  {author} {\bibfnamefont {M.-H.}\ \bibnamefont {Yung}}, \bibinfo {author}
  {\bibfnamefont {X.-Q.}\ \bibnamefont {Zhou}}, \bibinfo {author}
  {\bibfnamefont {P.~J.}\ \bibnamefont {Love}}, \bibinfo {author}
  {\bibfnamefont {A.}~\bibnamefont {Aspuru-Guzik}},\ and\ \bibinfo {author}
  {\bibfnamefont {J.~L.}\ \bibnamefont {O'Brien}},\ }\bibfield  {title}
  {\bibinfo {title} {A variational eigenvalue solver on a photonic quantum
  processor},\ }\href {https://doi.org/10.1038/ncomms5213} {\bibfield
  {journal} {\bibinfo  {journal} {Nature Communications}\ }\textbf {\bibinfo
  {volume} {5}},\ \bibinfo {pages} {4213} (\bibinfo {year} {2014})}\BibitemShut
  {NoStop}%
\bibitem [{\citenamefont {McClean}\ \emph {et~al.}(2016)\citenamefont
  {McClean}, \citenamefont {Romero}, \citenamefont {Babbush},\ and\
  \citenamefont {Aspuru-Guzik}}]{VarAlg1}%
  \BibitemOpen
  \bibfield  {author} {\bibinfo {author} {\bibfnamefont {J.~R.}\ \bibnamefont
  {McClean}}, \bibinfo {author} {\bibfnamefont {J.}~\bibnamefont {Romero}},
  \bibinfo {author} {\bibfnamefont {R.}~\bibnamefont {Babbush}},\ and\ \bibinfo
  {author} {\bibfnamefont {A.}~\bibnamefont {Aspuru-Guzik}},\ }\bibfield
  {title} {\bibinfo {title} {The theory of variational hybrid quantum-classical
  algorithms},\ }\href {https://doi.org/10.1088/1367-2630/18/2/023023}
  {\bibfield  {journal} {\bibinfo  {journal} {New Journal of Physics}\ }\textbf
  {\bibinfo {volume} {18}},\ \bibinfo {pages} {023023} (\bibinfo {year}
  {2016})}\BibitemShut {NoStop}%
\bibitem [{\citenamefont {{Farhi}}\ \emph {et~al.}(2000)\citenamefont
  {{Farhi}}, \citenamefont {{Goldstone}}, \citenamefont {{Gutmann}},\ and\
  \citenamefont {{Sipser}}}]{AdiabSim2}%
  \BibitemOpen
  \bibfield  {author} {\bibinfo {author} {\bibfnamefont {E.}~\bibnamefont
  {{Farhi}}}, \bibinfo {author} {\bibfnamefont {J.}~\bibnamefont
  {{Goldstone}}}, \bibinfo {author} {\bibfnamefont {S.}~\bibnamefont
  {{Gutmann}}},\ and\ \bibinfo {author} {\bibfnamefont {M.}~\bibnamefont
  {{Sipser}}},\ }\bibfield  {title} {\bibinfo {title} {{Quantum Computation by
  Adiabatic Evolution}},\ }\href@noop {} {\bibfield  {journal} {\bibinfo
  {journal} {arXiv e-prints}\ ,\ \bibinfo {eid} {quant-ph/0001106}} (\bibinfo
  {year} {2000})},\ \Eprint {https://arxiv.org/abs/quant-ph/0001106}
  {arXiv:quant-ph/0001106 [quant-ph]} \BibitemShut {NoStop}%
\bibitem [{\citenamefont {Born}\ and\ \citenamefont {Fock}(1928)}]{AdiabSim0}%
  \BibitemOpen
  \bibfield  {author} {\bibinfo {author} {\bibfnamefont {M.}~\bibnamefont
  {Born}}\ and\ \bibinfo {author} {\bibfnamefont {V.}~\bibnamefont {Fock}},\
  }\bibfield  {title} {\bibinfo {title} {Proof of the adiabatic theorem},\
  }\href {https://doi.org/10.1007/BF01343193} {\bibfield  {journal} {\bibinfo
  {journal} {Zeitschrift f{\"u}r Physik}\ }\textbf {\bibinfo {volume} {51}},\
  \bibinfo {pages} {165} (\bibinfo {year} {1928})}\BibitemShut {NoStop}%
\bibitem [{\citenamefont {Kato}(1950)}]{AdiabSim1}%
  \BibitemOpen
  \bibfield  {author} {\bibinfo {author} {\bibfnamefont {T.}~\bibnamefont
  {Kato}},\ }\bibfield  {title} {\bibinfo {title} {On the adiabatic theorem of
  quantum mechanics},\ }\href {https://doi.org/10.1143/JPSJ.5.435} {\bibfield
  {journal} {\bibinfo  {journal} {Journal of the Physical Society of Japan}\
  }\textbf {\bibinfo {volume} {5}},\ \bibinfo {pages} {435} (\bibinfo {year}
  {1950})}\BibitemShut {NoStop}%
\bibitem [{\citenamefont {Deutsch}(1991)}]{eth0}%
  \BibitemOpen
  \bibfield  {author} {\bibinfo {author} {\bibfnamefont {J.~M.}\ \bibnamefont
  {Deutsch}},\ }\bibfield  {title} {\bibinfo {title} {Quantum statistical
  mechanics in a closed system},\ }\href
  {https://doi.org/10.1103/PhysRevA.43.2046} {\bibfield  {journal} {\bibinfo
  {journal} {Phys. Rev. A}\ }\textbf {\bibinfo {volume} {43}},\ \bibinfo
  {pages} {2046} (\bibinfo {year} {1991})}\BibitemShut {NoStop}%
\bibitem [{\citenamefont {Srednicki}(1994)}]{eth1}%
  \BibitemOpen
  \bibfield  {author} {\bibinfo {author} {\bibfnamefont {M.}~\bibnamefont
  {Srednicki}},\ }\bibfield  {title} {\bibinfo {title} {Chaos and quantum
  thermalization},\ }\href {https://doi.org/10.1103/PhysRevE.50.888} {\bibfield
   {journal} {\bibinfo  {journal} {Phys. Rev. E}\ }\textbf {\bibinfo {volume}
  {50}},\ \bibinfo {pages} {888} (\bibinfo {year} {1994})}\BibitemShut
  {NoStop}%
\bibitem [{\citenamefont {Rigol}\ \emph {et~al.}(2008)\citenamefont {Rigol},
  \citenamefont {Dunjko},\ and\ \citenamefont {Olshanii}}]{eth2}%
  \BibitemOpen
  \bibfield  {author} {\bibinfo {author} {\bibfnamefont {M.}~\bibnamefont
  {Rigol}}, \bibinfo {author} {\bibfnamefont {V.}~\bibnamefont {Dunjko}},\ and\
  \bibinfo {author} {\bibfnamefont {M.}~\bibnamefont {Olshanii}},\ }\bibfield
  {title} {\bibinfo {title} {Thermalization and its mechanism for generic
  isolated quantum systems},\ }\href {https://doi.org/10.1038/nature06838}
  {\bibfield  {journal} {\bibinfo  {journal} {Nature}\ }\textbf {\bibinfo
  {volume} {452}},\ \bibinfo {pages} {854} (\bibinfo {year}
  {2008})}\BibitemShut {NoStop}%
\bibitem [{\citenamefont {Cassidy}\ \emph {et~al.}(2011)\citenamefont
  {Cassidy}, \citenamefont {Clark},\ and\ \citenamefont {Rigol}}]{eth3}%
  \BibitemOpen
  \bibfield  {author} {\bibinfo {author} {\bibfnamefont {A.~C.}\ \bibnamefont
  {Cassidy}}, \bibinfo {author} {\bibfnamefont {C.~W.}\ \bibnamefont {Clark}},\
  and\ \bibinfo {author} {\bibfnamefont {M.}~\bibnamefont {Rigol}},\ }\bibfield
   {title} {\bibinfo {title} {Generalized thermalization in an integrable
  lattice system},\ }\href {https://doi.org/10.1103/PhysRevLett.106.140405}
  {\bibfield  {journal} {\bibinfo  {journal} {Phys. Rev. Lett.}\ }\textbf
  {\bibinfo {volume} {106}},\ \bibinfo {pages} {140405} (\bibinfo {year}
  {2011})}\BibitemShut {NoStop}%
\bibitem [{\citenamefont {Rigol}\ and\ \citenamefont {Srednicki}(2012)}]{eth4}%
  \BibitemOpen
  \bibfield  {author} {\bibinfo {author} {\bibfnamefont {M.}~\bibnamefont
  {Rigol}}\ and\ \bibinfo {author} {\bibfnamefont {M.}~\bibnamefont
  {Srednicki}},\ }\bibfield  {title} {\bibinfo {title} {Alternatives to
  eigenstate thermalization},\ }\href
  {https://doi.org/10.1103/PhysRevLett.108.110601} {\bibfield  {journal}
  {\bibinfo  {journal} {Phys. Rev. Lett.}\ }\textbf {\bibinfo {volume} {108}},\
  \bibinfo {pages} {110601} (\bibinfo {year} {2012})}\BibitemShut {NoStop}%
\bibitem [{\citenamefont {M{\"u}ller}\ \emph {et~al.}(2015)\citenamefont
  {M{\"u}ller}, \citenamefont {Adlam}, \citenamefont {Masanes},\ and\
  \citenamefont {Wiebe}}]{eth4a}%
  \BibitemOpen
  \bibfield  {author} {\bibinfo {author} {\bibfnamefont {M.~P.}\ \bibnamefont
  {M{\"u}ller}}, \bibinfo {author} {\bibfnamefont {E.}~\bibnamefont {Adlam}},
  \bibinfo {author} {\bibfnamefont {L.}~\bibnamefont {Masanes}},\ and\ \bibinfo
  {author} {\bibfnamefont {N.}~\bibnamefont {Wiebe}},\ }\bibfield  {title}
  {\bibinfo {title} {Thermalization and canonical typicality in
  translation-invariant quantum lattice systems},\ }\href
  {https://doi.org/10.1007/s00220-015-2473-y} {\bibfield  {journal} {\bibinfo
  {journal} {Communications in Mathematical Physics}\ }\textbf {\bibinfo
  {volume} {340}},\ \bibinfo {pages} {499} (\bibinfo {year}
  {2015})}\BibitemShut {NoStop}%
\bibitem [{\citenamefont {Xu}\ \emph {et~al.}(2019)\citenamefont {Xu},
  \citenamefont {Li}, \citenamefont {Hsu}, \citenamefont {Swingle},\ and\
  \citenamefont {Das~Sarma}}]{eth5}%
  \BibitemOpen
  \bibfield  {author} {\bibinfo {author} {\bibfnamefont {S.}~\bibnamefont
  {Xu}}, \bibinfo {author} {\bibfnamefont {X.}~\bibnamefont {Li}}, \bibinfo
  {author} {\bibfnamefont {Y.-T.}\ \bibnamefont {Hsu}}, \bibinfo {author}
  {\bibfnamefont {B.}~\bibnamefont {Swingle}},\ and\ \bibinfo {author}
  {\bibfnamefont {S.}~\bibnamefont {Das~Sarma}},\ }\bibfield  {title} {\bibinfo
  {title} {Butterfly effect in interacting aubry-andre model: Thermalization,
  slow scrambling, and many-body localization},\ }\href
  {https://doi.org/10.1103/PhysRevResearch.1.032039} {\bibfield  {journal}
  {\bibinfo  {journal} {Phys. Rev. Research}\ }\textbf {\bibinfo {volume}
  {1}},\ \bibinfo {pages} {032039} (\bibinfo {year} {2019})}\BibitemShut
  {NoStop}%
\bibitem [{\citenamefont {Richter}\ \emph {et~al.}(2020)\citenamefont
  {Richter}, \citenamefont {Dymarsky}, \citenamefont {Steinigeweg},\ and\
  \citenamefont {Gemmer}}]{eth6}%
  \BibitemOpen
  \bibfield  {author} {\bibinfo {author} {\bibfnamefont {J.}~\bibnamefont
  {Richter}}, \bibinfo {author} {\bibfnamefont {A.}~\bibnamefont {Dymarsky}},
  \bibinfo {author} {\bibfnamefont {R.}~\bibnamefont {Steinigeweg}},\ and\
  \bibinfo {author} {\bibfnamefont {J.}~\bibnamefont {Gemmer}},\ }\bibfield
  {title} {\bibinfo {title} {Eigenstate thermalization hypothesis beyond
  standard indicators: Emergence of random-matrix behavior at small
  frequencies},\ }\href {https://doi.org/10.1103/PhysRevE.102.042127}
  {\bibfield  {journal} {\bibinfo  {journal} {Phys. Rev. E}\ }\textbf {\bibinfo
  {volume} {102}},\ \bibinfo {pages} {042127} (\bibinfo {year}
  {2020})}\BibitemShut {NoStop}%
\bibitem [{\citenamefont {Kuno}\ \emph {et~al.}(2020)\citenamefont {Kuno},
  \citenamefont {Mizoguchi},\ and\ \citenamefont {Hatsugai}}]{eth7}%
  \BibitemOpen
  \bibfield  {author} {\bibinfo {author} {\bibfnamefont {Y.}~\bibnamefont
  {Kuno}}, \bibinfo {author} {\bibfnamefont {T.}~\bibnamefont {Mizoguchi}},\
  and\ \bibinfo {author} {\bibfnamefont {Y.}~\bibnamefont {Hatsugai}},\
  }\bibfield  {title} {\bibinfo {title} {Flat band quantum scar},\ }\href
  {https://doi.org/10.1103/PhysRevB.102.241115} {\bibfield  {journal} {\bibinfo
   {journal} {Phys. Rev. B}\ }\textbf {\bibinfo {volume} {102}},\ \bibinfo
  {pages} {241115} (\bibinfo {year} {2020})}\BibitemShut {NoStop}%
\bibitem [{\citenamefont {Schuckert}\ and\ \citenamefont {Knap}(2020)}]{eth8}%
  \BibitemOpen
  \bibfield  {author} {\bibinfo {author} {\bibfnamefont {A.}~\bibnamefont
  {Schuckert}}\ and\ \bibinfo {author} {\bibfnamefont {M.}~\bibnamefont
  {Knap}},\ }\bibfield  {title} {\bibinfo {title} {Probing eigenstate
  thermalization in quantum simulators via fluctuation-dissipation relations},\
  }\href {https://doi.org/10.1103/PhysRevResearch.2.043315} {\bibfield
  {journal} {\bibinfo  {journal} {Phys. Rev. Research}\ }\textbf {\bibinfo
  {volume} {2}},\ \bibinfo {pages} {043315} (\bibinfo {year}
  {2020})}\BibitemShut {NoStop}%
\bibitem [{\citenamefont {{Cipolloni}}\ \emph {et~al.}(2020)\citenamefont
  {{Cipolloni}}, \citenamefont {{Erd{\H{o}}s}},\ and\ \citenamefont
  {{Schr{\"o}der}}}]{eth9}%
  \BibitemOpen
  \bibfield  {author} {\bibinfo {author} {\bibfnamefont {G.}~\bibnamefont
  {{Cipolloni}}}, \bibinfo {author} {\bibfnamefont {L.}~\bibnamefont
  {{Erd{\H{o}}s}}},\ and\ \bibinfo {author} {\bibfnamefont {D.}~\bibnamefont
  {{Schr{\"o}der}}},\ }\bibfield  {title} {\bibinfo {title} {{Eigenstate
  Thermalization Hypothesis for Wigner Matrices}},\ }\href@noop {} {\bibfield
  {journal} {\bibinfo  {journal} {arXiv e-prints}\ ,\ \bibinfo {eid}
  {arXiv:2012.13215}} (\bibinfo {year} {2020})},\ \Eprint
  {https://arxiv.org/abs/2012.13215} {arXiv:2012.13215 [math.PR]} \BibitemShut
  {NoStop}%
\bibitem [{\citenamefont {Noh}(2021)}]{eth10}%
  \BibitemOpen
  \bibfield  {author} {\bibinfo {author} {\bibfnamefont {J.~D.}\ \bibnamefont
  {Noh}},\ }\bibfield  {title} {\bibinfo {title} {Eigenstate thermalization
  hypothesis and eigenstate-to-eigenstate fluctuations},\ }\href
  {https://doi.org/10.1103/PhysRevE.103.012129} {\bibfield  {journal} {\bibinfo
   {journal} {Phys. Rev. E}\ }\textbf {\bibinfo {volume} {103}},\ \bibinfo
  {pages} {012129} (\bibinfo {year} {2021})}\BibitemShut {NoStop}%
\bibitem [{\citenamefont {{Huang}}(2021)}]{eth11}%
  \BibitemOpen
  \bibfield  {author} {\bibinfo {author} {\bibfnamefont {Y.}~\bibnamefont
  {{Huang}}},\ }\bibfield  {title} {\bibinfo {title} {{Finite-size scaling
  analysis of eigenstate thermalization}},\ }\href@noop {} {\bibfield
  {journal} {\bibinfo  {journal} {arXiv e-prints}\ ,\ \bibinfo {eid}
  {arXiv:2103.01539}} (\bibinfo {year} {2021})},\ \Eprint
  {https://arxiv.org/abs/2103.01539} {arXiv:2103.01539 [cond-mat.stat-mech]}
  \BibitemShut {NoStop}%
\bibitem [{\citenamefont {Nakerst}\ and\ \citenamefont {Haque}(2021)}]{eth12}%
  \BibitemOpen
  \bibfield  {author} {\bibinfo {author} {\bibfnamefont {G.}~\bibnamefont
  {Nakerst}}\ and\ \bibinfo {author} {\bibfnamefont {M.}~\bibnamefont
  {Haque}},\ }\bibfield  {title} {\bibinfo {title} {Eigenstate thermalization
  scaling in approaching the classical limit},\ }\href
  {https://doi.org/10.1103/PhysRevE.103.042109} {\bibfield  {journal} {\bibinfo
   {journal} {Phys. Rev. E}\ }\textbf {\bibinfo {volume} {103}},\ \bibinfo
  {pages} {042109} (\bibinfo {year} {2021})}\BibitemShut {NoStop}%
\bibitem [{\citenamefont {{Halataei}}(2021)}]{eth13}%
  \BibitemOpen
  \bibfield  {author} {\bibinfo {author} {\bibfnamefont {S.~M.~H.}\
  \bibnamefont {{Halataei}}},\ }\bibfield  {title} {\bibinfo {title} {{On
  eigenstate thermalization in the SYK chain model}},\ }\href@noop {}
  {\bibfield  {journal} {\bibinfo  {journal} {arXiv e-prints}\ ,\ \bibinfo
  {eid} {arXiv:2104.05291}} (\bibinfo {year} {2021})},\ \Eprint
  {https://arxiv.org/abs/2104.05291} {arXiv:2104.05291 [hep-th]} \BibitemShut
  {NoStop}%
\bibitem [{\citenamefont {Sch\"onle}\ \emph {et~al.}(2021)\citenamefont
  {Sch\"onle}, \citenamefont {Jansen}, \citenamefont {Heidrich-Meisner},\ and\
  \citenamefont {Vidmar}}]{eth14}%
  \BibitemOpen
  \bibfield  {author} {\bibinfo {author} {\bibfnamefont {C.}~\bibnamefont
  {Sch\"onle}}, \bibinfo {author} {\bibfnamefont {D.}~\bibnamefont {Jansen}},
  \bibinfo {author} {\bibfnamefont {F.}~\bibnamefont {Heidrich-Meisner}},\ and\
  \bibinfo {author} {\bibfnamefont {L.}~\bibnamefont {Vidmar}},\ }\bibfield
  {title} {\bibinfo {title} {Eigenstate thermalization hypothesis through the
  lens of autocorrelation functions},\ }\href
  {https://doi.org/10.1103/PhysRevB.103.235137} {\bibfield  {journal} {\bibinfo
   {journal} {Phys. Rev. B}\ }\textbf {\bibinfo {volume} {103}},\ \bibinfo
  {pages} {235137} (\bibinfo {year} {2021})}\BibitemShut {NoStop}%
\bibitem [{\citenamefont {{Fritzsch}}\ and\ \citenamefont
  {{Prosen}}(2021)}]{eth15}%
  \BibitemOpen
  \bibfield  {author} {\bibinfo {author} {\bibfnamefont {F.}~\bibnamefont
  {{Fritzsch}}}\ and\ \bibinfo {author} {\bibfnamefont {T.}~\bibnamefont
  {{Prosen}}},\ }\bibfield  {title} {\bibinfo {title} {{Eigenstate
  thermalization in dual-unitary quantum circuits: Asymptotics of spectral
  functions}},\ }\href {https://doi.org/10.1103/PhysRevE.103.062133} {\bibfield
   {journal} {\bibinfo  {journal} {\pre}\ }\textbf {\bibinfo {volume} {103}},\
  \bibinfo {eid} {062133} (\bibinfo {year} {2021})},\ \Eprint
  {https://arxiv.org/abs/2103.11694} {arXiv:2103.11694 [cond-mat.stat-mech]}
  \BibitemShut {NoStop}%
\bibitem [{\citenamefont {{Balachandran}}\ \emph {et~al.}(2021)\citenamefont
  {{Balachandran}}, \citenamefont {{Benenti}}, \citenamefont {{Casati}},\ and\
  \citenamefont {{Poletti}}}]{eth16}%
  \BibitemOpen
  \bibfield  {author} {\bibinfo {author} {\bibfnamefont {V.}~\bibnamefont
  {{Balachandran}}}, \bibinfo {author} {\bibfnamefont {G.}~\bibnamefont
  {{Benenti}}}, \bibinfo {author} {\bibfnamefont {G.}~\bibnamefont
  {{Casati}}},\ and\ \bibinfo {author} {\bibfnamefont {D.}~\bibnamefont
  {{Poletti}}},\ }\bibfield  {title} {\bibinfo {title} {{From ETH to algebraic
  relaxation of OTOCs in systems with conserved quantities}},\ }\href@noop {}
  {\bibfield  {journal} {\bibinfo  {journal} {arXiv e-prints}\ ,\ \bibinfo
  {eid} {arXiv:2106.00234}} (\bibinfo {year} {2021})},\ \Eprint
  {https://arxiv.org/abs/2106.00234} {arXiv:2106.00234 [cond-mat.stat-mech]}
  \BibitemShut {NoStop}%
\bibitem [{\citenamefont {{Decker}}\ \emph {et~al.}(2021)\citenamefont
  {{Decker}}, \citenamefont {{Kennes}},\ and\ \citenamefont
  {{Karrasch}}}]{eth17}%
  \BibitemOpen
  \bibfield  {author} {\bibinfo {author} {\bibfnamefont {K.~S.~C.}\
  \bibnamefont {{Decker}}}, \bibinfo {author} {\bibfnamefont {D.~M.}\
  \bibnamefont {{Kennes}}},\ and\ \bibinfo {author} {\bibfnamefont
  {C.}~\bibnamefont {{Karrasch}}},\ }\bibfield  {title} {\bibinfo {title}
  {{Many-body localization and the area law in two dimensions}},\ }\href@noop
  {} {\bibfield  {journal} {\bibinfo  {journal} {arXiv e-prints}\ ,\ \bibinfo
  {eid} {arXiv:2106.12861}} (\bibinfo {year} {2021})},\ \Eprint
  {https://arxiv.org/abs/2106.12861} {arXiv:2106.12861 [cond-mat.dis-nn]}
  \BibitemShut {NoStop}%
\bibitem [{\citenamefont {{De Palma}}\ and\ \citenamefont
  {{Rouz{\'e}}}(2021)}]{eth18}%
  \BibitemOpen
  \bibfield  {author} {\bibinfo {author} {\bibfnamefont {G.}~\bibnamefont {{De
  Palma}}}\ and\ \bibinfo {author} {\bibfnamefont {C.}~\bibnamefont
  {{Rouz{\'e}}}},\ }\bibfield  {title} {\bibinfo {title} {{Quantum
  concentration inequalities}},\ }\href@noop {} {\bibfield  {journal} {\bibinfo
   {journal} {arXiv e-prints}\ ,\ \bibinfo {eid} {arXiv:2106.15819}} (\bibinfo
  {year} {2021})},\ \Eprint {https://arxiv.org/abs/2106.15819}
  {arXiv:2106.15819 [quant-ph]} \BibitemShut {NoStop}%
\bibitem [{\citenamefont {{Khudorozhkov}}\ \emph {et~al.}(2021)\citenamefont
  {{Khudorozhkov}}, \citenamefont {{Tiwari}}, \citenamefont {{Chamon}},\ and\
  \citenamefont {{Neupert}}}]{eth19}%
  \BibitemOpen
  \bibfield  {author} {\bibinfo {author} {\bibfnamefont {A.}~\bibnamefont
  {{Khudorozhkov}}}, \bibinfo {author} {\bibfnamefont {A.}~\bibnamefont
  {{Tiwari}}}, \bibinfo {author} {\bibfnamefont {C.}~\bibnamefont {{Chamon}}},\
  and\ \bibinfo {author} {\bibfnamefont {T.}~\bibnamefont {{Neupert}}},\
  }\bibfield  {title} {\bibinfo {title} {{Hilbert space fragmentation in a 2D
  quantum spin system with subsystem symmetries}},\ }\href@noop {} {\bibfield
  {journal} {\bibinfo  {journal} {arXiv e-prints}\ ,\ \bibinfo {eid}
  {arXiv:2107.09690}} (\bibinfo {year} {2021})},\ \Eprint
  {https://arxiv.org/abs/2107.09690} {arXiv:2107.09690 [cond-mat.str-el]}
  \BibitemShut {NoStop}%
\bibitem [{\citenamefont {{Mukherjee}}\ \emph {et~al.}(2021)\citenamefont
  {{Mukherjee}}, \citenamefont {{Cai}},\ and\ \citenamefont {{Liu}}}]{eth20}%
  \BibitemOpen
  \bibfield  {author} {\bibinfo {author} {\bibfnamefont {B.}~\bibnamefont
  {{Mukherjee}}}, \bibinfo {author} {\bibfnamefont {Z.}~\bibnamefont {{Cai}}},\
  and\ \bibinfo {author} {\bibfnamefont {W.~V.}\ \bibnamefont {{Liu}}},\
  }\bibfield  {title} {\bibinfo {title} {{Constraint-induced breaking and
  restoration of ergodicity in spin-1 PXP models}},\ }\href
  {https://doi.org/10.1103/PhysRevResearch.3.033201} {\bibfield  {journal}
  {\bibinfo  {journal} {Physical Review Research}\ }\textbf {\bibinfo {volume}
  {3}},\ \bibinfo {eid} {033201} (\bibinfo {year} {2021})},\ \Eprint
  {https://arxiv.org/abs/2104.00699} {arXiv:2104.00699 [quant-ph]} \BibitemShut
  {NoStop}%
\bibitem [{\citenamefont {Whittaker}(1915)}]{Nyquist0}%
  \BibitemOpen
  \bibfield  {author} {\bibinfo {author} {\bibfnamefont {E.}~\bibnamefont
  {Whittaker}},\ }\bibfield  {title} {\bibinfo {title} {On the functions which
  are represented by the expansions of the interpolation-theory},\ }\href
  {https://doi.org/10.1017/S0370164600017806} {\bibfield  {journal} {\bibinfo
  {journal} {Proceedings of the Royal Society of Edinburgh}\ }\textbf {\bibinfo
  {volume} {35}},\ \bibinfo {pages} {181} (\bibinfo {year} {1915})}\BibitemShut
  {NoStop}%
\bibitem [{\citenamefont {Nyquist}(1928)}]{Nyquist1}%
  \BibitemOpen
  \bibfield  {author} {\bibinfo {author} {\bibfnamefont {H.}~\bibnamefont
  {Nyquist}},\ }\bibfield  {title} {\bibinfo {title} {Certain topics in
  telegraph transmission theory},\ }\href
  {https://doi.org/10.1109/T-AIEE.1928.5055024} {\bibfield  {journal} {\bibinfo
   {journal} {Transactions of the American Institute of Electrical Engineers}\
  }\textbf {\bibinfo {volume} {47}},\ \bibinfo {pages} {617} (\bibinfo {year}
  {1928})}\BibitemShut {NoStop}%
\bibitem [{\citenamefont {Shannon}(1949)}]{Nyquist2}%
  \BibitemOpen
  \bibfield  {author} {\bibinfo {author} {\bibfnamefont {C.}~\bibnamefont
  {Shannon}},\ }\bibfield  {title} {\bibinfo {title} {Communication in the
  presence of noise},\ }\href {https://doi.org/10.1109/JRPROC.1949.232969}
  {\bibfield  {journal} {\bibinfo  {journal} {Proceedings of the IRE}\ }\textbf
  {\bibinfo {volume} {37}},\ \bibinfo {pages} {10} (\bibinfo {year}
  {1949})}\BibitemShut {NoStop}%
\bibitem [{\citenamefont {Efron}(1979)}]{bootstrapI}%
  \BibitemOpen
  \bibfield  {author} {\bibinfo {author} {\bibfnamefont {B.}~\bibnamefont
  {Efron}},\ }\bibfield  {title} {\bibinfo {title} {Bootstrap methods: Another
  look at the jackknife},\ }\href {https://doi.org/10.1214/aos/1176344552}
  {\bibfield  {journal} {\bibinfo  {journal} {Ann. Statist.}\ }\textbf
  {\bibinfo {volume} {7}},\ \bibinfo {pages} {1} (\bibinfo {year}
  {1979})}\BibitemShut {NoStop}%
\bibitem [{\citenamefont {Kunsch}(1989)}]{bootstrapII}%
  \BibitemOpen
  \bibfield  {author} {\bibinfo {author} {\bibfnamefont {H.~R.}\ \bibnamefont
  {Kunsch}},\ }\bibfield  {title} {\bibinfo {title} {The jackknife and the
  bootstrap for general stationary observations},\ }\href
  {https://doi.org/10.1214/aos/1176347265} {\bibfield  {journal} {\bibinfo
  {journal} {Ann. Statist.}\ }\textbf {\bibinfo {volume} {17}},\ \bibinfo
  {pages} {1217} (\bibinfo {year} {1989})}\BibitemShut {NoStop}%
\bibitem [{\citenamefont {Politis}\ and\ \citenamefont
  {Romano}(1994)}]{bootstrapIII}%
  \BibitemOpen
  \bibfield  {author} {\bibinfo {author} {\bibfnamefont {D.~N.}\ \bibnamefont
  {Politis}}\ and\ \bibinfo {author} {\bibfnamefont {J.~P.}\ \bibnamefont
  {Romano}},\ }\bibfield  {title} {\bibinfo {title} {The stationary
  bootstrap},\ }\href {https://doi.org/10.1080/01621459.1994.10476870}
  {\bibfield  {journal} {\bibinfo  {journal} {Journal of the American
  Statistical Association}\ }\textbf {\bibinfo {volume} {89}},\ \bibinfo
  {pages} {1303} (\bibinfo {year} {1994})}\BibitemShut {NoStop}%
\bibitem [{\citenamefont {Kroese}\ \emph {et~al.}(2014)\citenamefont {Kroese},
  \citenamefont {Brereton}, \citenamefont {Taimre},\ and\ \citenamefont
  {Botev}}]{MonteTech0}%
  \BibitemOpen
  \bibfield  {author} {\bibinfo {author} {\bibfnamefont {D.~P.}\ \bibnamefont
  {Kroese}}, \bibinfo {author} {\bibfnamefont {T.}~\bibnamefont {Brereton}},
  \bibinfo {author} {\bibfnamefont {T.}~\bibnamefont {Taimre}},\ and\ \bibinfo
  {author} {\bibfnamefont {Z.~I.}\ \bibnamefont {Botev}},\ }\bibfield  {title}
  {\bibinfo {title} {Why the monte carlo method is so important today},\ }\href
  {https://doi.org/https://doi.org/10.1002/wics.1314} {\bibfield  {journal}
  {\bibinfo  {journal} {WIREs Computational Statistics}\ }\textbf {\bibinfo
  {volume} {6}},\ \bibinfo {pages} {386} (\bibinfo {year} {2014})}\BibitemShut
  {NoStop}%
\bibitem [{\citenamefont {Metropolis}\ \emph {et~al.}(1953)\citenamefont
  {Metropolis}, \citenamefont {Rosenbluth}, \citenamefont {Rosenbluth},
  \citenamefont {Teller},\ and\ \citenamefont {Teller}}]{Metro0}%
  \BibitemOpen
  \bibfield  {author} {\bibinfo {author} {\bibfnamefont {N.}~\bibnamefont
  {Metropolis}}, \bibinfo {author} {\bibfnamefont {A.~W.}\ \bibnamefont
  {Rosenbluth}}, \bibinfo {author} {\bibfnamefont {M.~N.}\ \bibnamefont
  {Rosenbluth}}, \bibinfo {author} {\bibfnamefont {A.~H.}\ \bibnamefont
  {Teller}},\ and\ \bibinfo {author} {\bibfnamefont {E.}~\bibnamefont
  {Teller}},\ }\bibfield  {title} {\bibinfo {title} {Equation of state
  calculations by fast computing machines},\ }\href
  {https://doi.org/10.1063/1.1699114} {\bibfield  {journal} {\bibinfo
  {journal} {The Journal of Chemical Physics}\ }\textbf {\bibinfo {volume}
  {21}},\ \bibinfo {pages} {1087} (\bibinfo {year} {1953})}\BibitemShut
  {NoStop}%
\bibitem [{\citenamefont {Hastings}(1970)}]{Metro1}%
  \BibitemOpen
  \bibfield  {author} {\bibinfo {author} {\bibfnamefont {W.~K.}\ \bibnamefont
  {Hastings}},\ }\bibfield  {title} {\bibinfo {title} {{Monte Carlo sampling
  methods using Markov chains and their applications}},\ }\href
  {https://doi.org/10.1093/biomet/57.1.97} {\bibfield  {journal} {\bibinfo
  {journal} {Biometrika}\ }\textbf {\bibinfo {volume} {57}},\ \bibinfo {pages}
  {97} (\bibinfo {year} {1970})}\BibitemShut {NoStop}%
\bibitem [{\citenamefont {Gilks}\ and\ \citenamefont {Wild}(1992)}]{Metro2}%
  \BibitemOpen
  \bibfield  {author} {\bibinfo {author} {\bibfnamefont {W.~R.}\ \bibnamefont
  {Gilks}}\ and\ \bibinfo {author} {\bibfnamefont {P.}~\bibnamefont {Wild}},\
  }\bibfield  {title} {\bibinfo {title} {Adaptive rejection sampling for gibbs
  sampling},\ }\href {http://www.jstor.org/stable/2347565} {\bibfield
  {journal} {\bibinfo  {journal}
  {\href{http://www.jstor.org/stable/2347565}{Journal of the Royal Statistical
  Society. Series C (Applied Statistics)}}\ }\textbf {\bibinfo {volume} {41}},\
  \bibinfo {pages} {337} (\bibinfo {year} {1992})}\BibitemShut {NoStop}%
\bibitem [{\citenamefont {Hill}\ and\ \citenamefont {Spall}(2019)}]{Metro3}%
  \BibitemOpen
  \bibfield  {author} {\bibinfo {author} {\bibfnamefont {S.~D.}\ \bibnamefont
  {Hill}}\ and\ \bibinfo {author} {\bibfnamefont {J.~C.}\ \bibnamefont
  {Spall}},\ }\bibfield  {title} {\bibinfo {title} {Stationarity and
  convergence of the metropolis-hastings algorithm: Insights into theoretical
  aspects},\ }\href {https://doi.org/10.1109/MCS.2018.2876959} {\bibfield
  {journal} {\bibinfo  {journal} {IEEE Control Systems Magazine}\ }\textbf
  {\bibinfo {volume} {39}},\ \bibinfo {pages} {56} (\bibinfo {year}
  {2019})}\BibitemShut {NoStop}%
\bibitem [{\citenamefont {Caflisch}(1998)}]{MonteCarlo0}%
  \BibitemOpen
  \bibfield  {author} {\bibinfo {author} {\bibfnamefont {R.~E.}\ \bibnamefont
  {Caflisch}},\ }\bibfield  {title} {\bibinfo {title} {Monte carlo and
  quasi-monte carlo methods},\ }\href
  {https://doi.org/10.1017/S0962492900002804} {\bibfield  {journal} {\bibinfo
  {journal} {Acta Numerica}\ }\textbf {\bibinfo {volume} {7}},\ \bibinfo
  {pages} {1} (\bibinfo {year} {1998})}\BibitemShut {NoStop}%
\bibitem [{\citenamefont {Cobham}(1965)}]{cobhamSim}%
  \BibitemOpen
  \bibfield  {author} {\bibinfo {author} {\bibfnamefont {A.}~\bibnamefont
  {Cobham}},\ }\bibfield  {title} {\bibinfo {title} {The intrinsic
  computational difficulty of functions},\ }\href
  {https://doi.org/10.2307/2270886} {\bibfield  {journal} {\bibinfo  {journal}
  {North-Holland Publishing}\ ,\ \bibinfo {pages} {24}} (\bibinfo {year}
  {1965})}\BibitemShut {NoStop}%
\bibitem [{\citenamefont {Lloyd}(1996)}]{complexTwo}%
  \BibitemOpen
  \bibfield  {author} {\bibinfo {author} {\bibfnamefont {S.}~\bibnamefont
  {Lloyd}},\ }\bibfield  {title} {\bibinfo {title} {Universal quantum
  simulators},\ }\href {https://doi.org/10.1126/science.273.5278.1073}
  {\bibfield  {journal} {\bibinfo  {journal} {Science}\ }\textbf {\bibinfo
  {volume} {273}},\ \bibinfo {pages} {1073} (\bibinfo {year}
  {1996})}\BibitemShut {NoStop}%
\bibitem [{\citenamefont {{Nagaj}}(2010)}]{completeIV}%
  \BibitemOpen
  \bibfield  {author} {\bibinfo {author} {\bibfnamefont {D.}~\bibnamefont
  {{Nagaj}}},\ }\bibfield  {title} {\bibinfo {title} {{Fast universal quantum
  computation with railroad-switch local Hamiltonians}},\ }\href
  {https://doi.org/10.1063/1.3384661} {\bibfield  {journal} {\bibinfo
  {journal} {Journal of Mathematical Physics}\ }\textbf {\bibinfo {volume}
  {51}},\ \bibinfo {pages} {062201} (\bibinfo {year} {2010})},\ \Eprint
  {https://arxiv.org/abs/0908.4219} {arXiv:0908.4219 [quant-ph]} \BibitemShut
  {NoStop}%
\bibitem [{\citenamefont {{Berry}}\ \emph {et~al.}(2015)\citenamefont
  {{Berry}}, \citenamefont {{Childs}},\ and\ \citenamefont
  {{Kothari}}}]{completeI}%
  \BibitemOpen
  \bibfield  {author} {\bibinfo {author} {\bibfnamefont {D.~W.}\ \bibnamefont
  {{Berry}}}, \bibinfo {author} {\bibfnamefont {A.~M.}\ \bibnamefont
  {{Childs}}},\ and\ \bibinfo {author} {\bibfnamefont {R.}~\bibnamefont
  {{Kothari}}},\ }\bibfield  {title} {\bibinfo {title} {{Hamiltonian simulation
  with nearly optimal dependence on all parameters}},\ }\href@noop {}
  {\bibfield  {journal} {\bibinfo  {journal} {arXiv e-prints}\ ,\ \bibinfo
  {eid} {arXiv:1501.01715}} (\bibinfo {year} {2015})},\ \Eprint
  {https://arxiv.org/abs/1501.01715} {arXiv:1501.01715 [quant-ph]} \BibitemShut
  {NoStop}%
\bibitem [{\citenamefont {{Hao Low}}\ and\ \citenamefont
  {{Chuang}}(2016)}]{completeII}%
  \BibitemOpen
  \bibfield  {author} {\bibinfo {author} {\bibfnamefont {G.}~\bibnamefont {{Hao
  Low}}}\ and\ \bibinfo {author} {\bibfnamefont {I.~L.}\ \bibnamefont
  {{Chuang}}},\ }\bibfield  {title} {\bibinfo {title} {{Hamiltonian Simulation
  by Qubitization}},\ }\href@noop {} {\bibfield  {journal} {\bibinfo  {journal}
  {arXiv e-prints}\ ,\ \bibinfo {eid} {arXiv:1610.06546}} (\bibinfo {year}
  {2016})},\ \Eprint {https://arxiv.org/abs/1610.06546} {arXiv:1610.06546
  [quant-ph]} \BibitemShut {NoStop}%
\bibitem [{\citenamefont {{Haah}}\ \emph {et~al.}(2018)\citenamefont {{Haah}},
  \citenamefont {{Hastings}}, \citenamefont {{Kothari}},\ and\ \citenamefont
  {{Hao Low}}}]{interI}%
  \BibitemOpen
  \bibfield  {author} {\bibinfo {author} {\bibfnamefont {J.}~\bibnamefont
  {{Haah}}}, \bibinfo {author} {\bibfnamefont {M.~B.}\ \bibnamefont
  {{Hastings}}}, \bibinfo {author} {\bibfnamefont {R.}~\bibnamefont
  {{Kothari}}},\ and\ \bibinfo {author} {\bibfnamefont {G.}~\bibnamefont {{Hao
  Low}}},\ }\bibfield  {title} {\bibinfo {title} {{Quantum algorithm for
  simulating real time evolution of lattice Hamiltonians}},\ }\href@noop {}
  {\bibfield  {journal} {\bibinfo  {journal} {arXiv e-prints}\ ,\ \bibinfo
  {eid} {arXiv:1801.03922}} (\bibinfo {year} {2018})},\ \Eprint
  {https://arxiv.org/abs/1801.03922} {arXiv:1801.03922 [quant-ph]} \BibitemShut
  {NoStop}%
\end{thebibliography}%

\end{document}